# Demonstrating Immersive Media Delivery on 5G Broadcast and Multicast Testing Networks

De Mi, Joe Eyles, Tero Jokela, Swen Petersen, Roman Odarchenko, Ece Öztürk, Duy-Kha Chau, Tuan Tran, Rory Turnbull, Heikki Kokkinen, Baruch Altman, Menno Bot, Darko Ratkaj, Olaf Renner, David Gomez-Barquero and Jordi Joan Gimenez

***Abstract* — This work presents eight demonstrators and one showcase developed within the 5G-Xcast project. They experimentally demonstrate and validate key technical enablers for the future of media delivery, associated with multicast and broadcast communication capabilities in 5th Generation (5G). In 5G-Xcast, three existing testbeds: IRT in Munich (Germany), 5GIC in Surrey (UK), and TUAS in Turku (Finland), have been developed into 5G broadcast and multicast testing networks, which enables us to demonstrate our vision of a converged 5G infrastructure with fixed and mobile accesses and terrestrial broadcast, delivering immersive audio-visual media content. Built upon the improved testing networks, the demonstrators and showcase developed in 5G-Xcast show the impact of the technology developed in the project. Our demonstrations predominantly cover use cases belonging to two verticals: Media & Entertainment and Public Warning, which are future 5G scenarios relevant to multicast and broadcast delivery. In this paper, we present the development of these demonstrators, the showcase, and the testbeds. We also provide key findings from the experiments and demonstrations, which not only validate the technical solutions developed in the project, but also illustrate the potential technical impact of these solutions for broadcasters, content providers, operators, and other industries interested in the future immersive media delivery.**

## I. Introduction

The 5th Generation (5G) of mobile communication promises unprecedented capacity, performance and flexibility to support a vast array of services. One of the most prominent nowadays, which deserves a lot of attention is the delivery of future media [1][2]. Applications in the context of future media delivery include ultra-high-definition television services [3], vehicular communications [4], object-based content delivery, and immersive audio-visual media services for popular live events with potentially millions of concurrent audiences [5][6]. To deliver these new audio-visual media services (and additionally including consumer interactivity) brings big challenges to cellular networks based solely on unicast mode, in particular when there is simultaneous demand for the same piece of content by a large number of users. Point-to-Multipoint (PTM) transmission such as broadcast and multicast can be a more efficient delivery mechanism compared to unicast whenever a service or an application requires the same content to be simultaneously delivered to multiple users or devices, even when users are concentrated in one or a small number of cells and are interested in the same content [7]. Therefore, multicast and broadcast communication capabilities are essential features for 5G applications in the future of media delivery.

The current 5G standardization by the 3rd Generation Partnership Project (3GPP) has focused so far on unicast, even though 3GPP has enhanced in Release-16 the broadcast mode of the 4th Generation (4G) Long Term Evolution (LTE) set of specifications [8][9], the enhanced Multimedia Broadcast Multicast Service (eMBMS). Release-17 is addressing multicast and broadcast features for the 5G system architecture and the radio access network with 5G New Radio (NR) [10].

The 5G Infrastructure Public Private Partnership (5G-PPP) phase-II project 5G-Xcast [11] has designed, assessed and

This work was supported in part by the European Commission under the 5GPPP project 5G-Xcast (H2020-ICT-2016-2 call, grant number 761498). The views expressed in this contribution are those of the authors and do not necessarily represent the project.

We thank the following individuals for their significant contributions in helping the development of the demonstrators and testbeds as shown in this work: Sam Hurst, Andrew Murphy, Richard Bradbury (BBC R&D, UK); Juha Kalliovaara, Jarkko Paavola (Turku University of Applied Sciences, Finland); Pei Xiao (University of Surrey, UK); Steve Appleby (BT, UK); Peter Sanders (One2Many, The Netherlands); Yann Moreaux (Expway-ENENSYS, France); Yann Begassat, Alain-Pierre Brunel, Maël Boutin (Broadpeak, France); Volker Pauli, Waqar Zia (Nomor Research GmbH, Germany); Ekkehard Lang (Nokia Bell Labs, Germany).

D. Mi is with University of Surrey, Guildford, GU2 7XH, United Kingdom (e-mail: d.mi@surrey.ac.uk).

J. Eyles is with BBC R&D, London, W12 7TQ, United Kingdom (email: joe.eyles@bbc.co.uk).

T. Jokela is with Turku University of Applied Sciences, Turku, 20520, Finland (e-mail: tero.jokela@turkuamk.fi).

S. Petersen and J.J. Gimenez are with Institut für Rundfunktechnik GmbH, Munich, 80939, Germany (e-mail: {petersen, jordi.gimenez}@irt.de).

R. Odarchenko is with BundlesLab Kft, Budapest, 1117, Hungary (e-mail: roman@bundleslab.com).

E. Öztürk is with Nomor Research GmbH, Munich, 81541, Germany (e-mail: oeztuerk@nomor.de).

D.K. Chau is with Broadpeak, Rennes, 35510, France (email: duy-kha.chau@broadpeak.tv).

T. Tran is with Enensys, Paris, 75007, France (e-mail: tuan.tran@enensys.com).

R. Turnbull is with BT, London, United Kingdom (e-mail: rory.turnbull@bt.com).

H. Kokkinen is with Fairspectrum OY, Espoo, 02150, Finland (e-mail: heikki.kokkinen@fairspectrum.com).

B. Altman is with LiveU Ltd, 44422111, Israel (e-mail: baruch@liveu.tv).

M. Bot is with One2Many, Deventer, 7411CL, The Netherlands (e-mail: menno.bot@one2many.eu).

D. Ratkaj is with European Broadcasting Union, Switzerland (e-mail: ratkaj@ebu.ch).

O. Renner is with Nokia Bell Labs, Munich, 81541, Germany (e-mail: olaf.renner@nokia.com).

D. Gomez-Barquero is with Universitat Politecnica de Valencia, Valencia, 46022, Spain (e-mail: dagobar@iteam.upv.es).

demonstrated a conceptually novel and forward-looking 5G network architecture for large scale immersive media delivery. The intention here is to include broadcast/multicast as a critical technology element in the second phase of 5G specifications and as a complement to unicast. Considering the forward compatibility of 5G NR, the design principle of 5G PTM is to ensure: 1) a seamless switching between unicast and broadcast/multicast [12]; 2) a smooth introduction of future PTM services and features with low impact on the architectures of both the network and user equipment (UE) [13][14]. To this end, it is required to first define relevant use cases for broadcast/multicast, for which technical solutions should then be designed, experimentally validated and finally demonstrated. Recall that audio-visual media services can generate considerably large volumes of data traffic on networks, where the data demand could be unevenly distributed over time and geographical areas. The target areas vary in size from small regions to the entirety of multiple countries. At the same time, Quality of Experience (QoE) is strongly dependent on maintaining sustained minimum data rates and low latencies to every user [15][16], regardless of the total number of concurrent users. This is particularly challenging for very popular live content (e.g. sports) or unpredictable events (e.g. breaking news) that tend to cause large traffic spikes. Hence, in 5G-Xcast, two vertical sectors are focused on, namely Media and Entertainment (M&E) and Public Warning (PW), and we further identify three use cases for demonstration purpose, as follows:

### *A. Media and Entertainment Use Cases*

In M&E use cases, broadcast and multicast transmissions facilitate the distribution of audio-visual media content and services, particularly to cover popular live events for a very large number of concurrent users [17][18].

*Hybrid Broadcast Service (HBS)*: In this use case, users have access to any combination of linear and non-linear audio-visual content in addition to social media. The content is diverse and includes multiple media types such as video, audio, text, and data, possibly coming from various sources, e.g., different content providers. Audio-visual media services can be personalized and combined with other functionalities, such as social media, location-based features, interactivity, interpersonal communications, and more. Access to content and media services is enabled on different user devices and in different environments. Content and services may be delivered over a combination of several networks and types of network simultaneously. Continuity of the users' experience should be preserved when switching between different access networks, possibly operated by different operators. The population of concurrent users are potentially very large and may substantially change over short periods of time [17].

*Object-based Media (OBM)*: Under an OBM approach [19], the programme is captured in a conventional way but stored as a set of its component parts, be they audio, video, captions or other material along with detailed metadata that describes how these should be assembled. These component parts are then delivered separately and rendered on the device in a way that takes account of the capabilities of the device, the environment and the user's preferences. The OBM use case goes beyond the traditional approach for delivering media services. The traditional method of programme making is to capture audio and video content, edit it down to produce a single linear audio/video stream and distribute or broadcast this to everyone. However, this may result in compromises when it is being viewed on certain devices in certain environments. For example, sizes of captions or audio mixes that are suitable for a large screen TV in the living room environment may not be optimal for consumption on a smartphone on the bus. The OBM concept is similar to the approach taken with a web page, which is rendered differently in the browser depending on whether it is viewed on a mobile device or personal computer. This rendering approach also opens the door to new content experiences and more immersive forms of experience such as virtual and augmented reality and 360-degree video.

### *B. Public Warning use Case*

A key requirement for PW applications is the secure and reliable delivery of alert messages to the general population in emergency situations. Broadcast and multicast capabilities can ensure that PW messages reach a large number of users simultaneously without causing network congestion or even significantly increasing the traffic load [17].

*Multimedia Public Warning Alert*: This use case concerns the delivery of multimedia PW message to a selected area. More contextual information can be added, which would currently be difficult to squeeze into the limited amount of text an alert typically has. This additional information could include: a geographic map of area of the alert, a recommended route, recommended actions, and alerts for multiple languages. These may be provided as picture or video objects, in different object types or resolution. Users can move between being outdoors or indoors. They would receive the PW message when within the area of an active alert. Accessibility for visually-impaired or hearing-impaired users could be greatly improved by providing, for example, audio for visually-impaired users, video with sign language for hearing-impaired users, or text for those users who do not need or want audio or video.

To experimentally demonstrate and validate the 5G-Xcast technical solutions targeting these use cases, we utilize three testbeds, including two urban city testbeds at the Institute für Rundfunktechnik (IRT) in Munich (Germany) and Turku University of Applied Sciences (TUAS) in Turku (Finland), and a university campus 5G-compatible testbed: the 5G Innovation Centre (5GIC) in Surrey (UK). The three testbeds are further extended in 5G-Xcast to form 5G testing networks, equipped with broadcast and multicast communication capabilities. Together with the 5G-Xcast partners we have built eight demonstrators and one showcase based on the extended testing networks. These present a showcase of the immersive media delivery in 5G. In the following sections, we will first describe the development of three testbeds, then elaborate on the developed showcase and demonstrators, and finally discuss the potential of the demonstrated 5G-Xcast technology solutions on broadcasting society and industry. Note that more details and results of the showcase and demonstrators can be found in 5G-Xcast public deliverables [20][21].

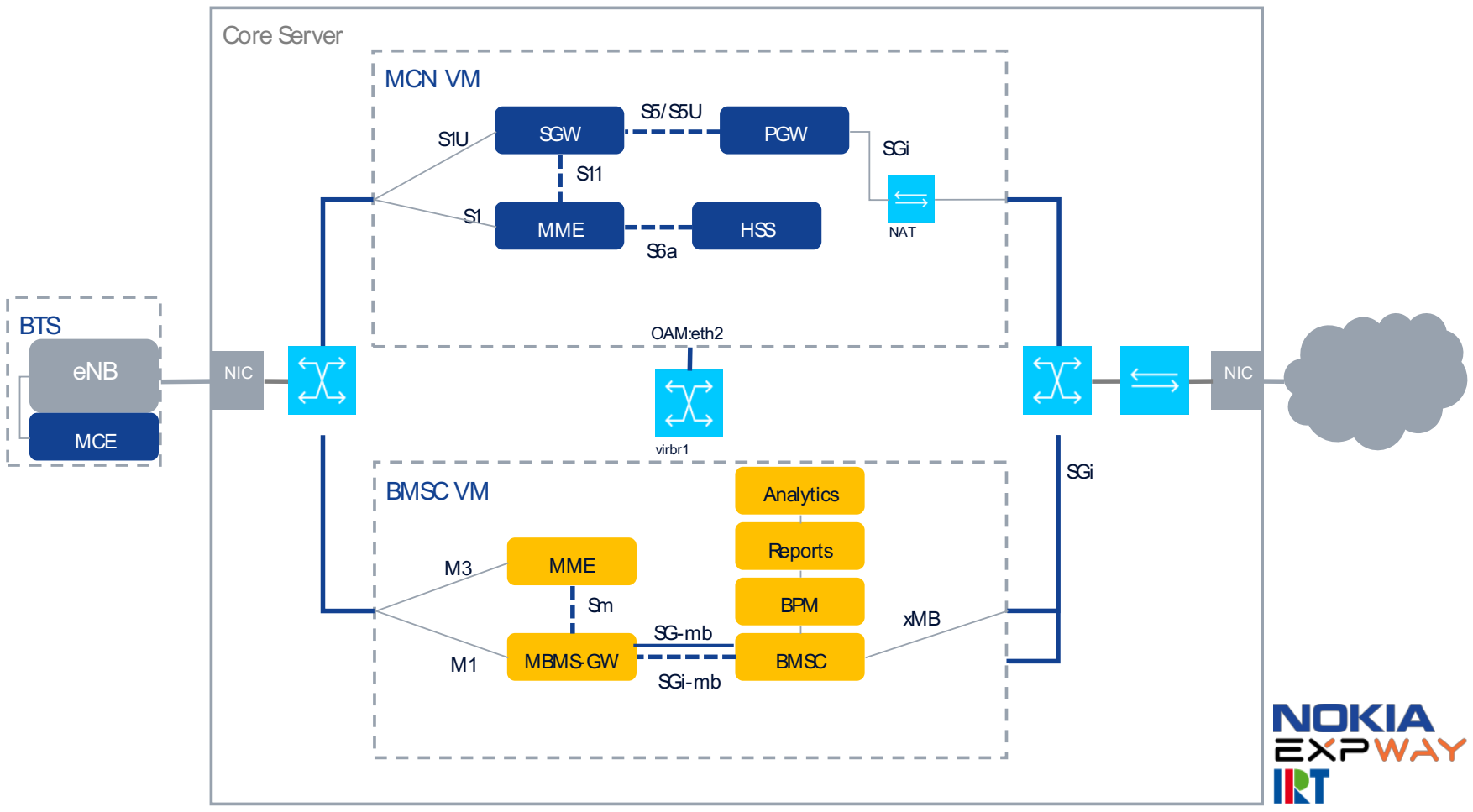

Fig. 1. Update of the IRT Munich testbed and integration of MooD.

## II. Development of 5G Broadcast and Multicast Testing Networks

Facilities for experimental purposes have been created in recent years to test technologies for a number of application areas currently targeted in 5G. The three testbeds used in the 5G-Xcast also incorporate fixed network capabilities to allow the testing of convergence scenarios between mobile, broadcast and fixed networks [22]. The testbeds and their development are described in the following subsections.

### *A. IRT Munich Testbed*

The original test facilities in Munich implemented LTE Release-11 eMBMS with a base station serving four sites via remote radio heads (RRH) ensuring synchronization for single frequency network (SFN) operation. The sites are located in the area surrounding Munich, covering a wide part of its urban area. The inter-site distances are in the range between 1.8 and 19.8 km, which permits configuring the Multicast Broadcast Single Frequency Network (MBSFN) with cyclic prefix (CP) 16.67 µs, providing enough coverage for testing purposes. The sites are operated in LTE band 28 (Frequency Division Duplexing - FDD, 706 to 716 MHz uplink, 761 to 771 MHz downlink). The transmit power of the network is 400 W effective isotropic radiated power (EIRP) per site/antenna [23].

The system supports the allocation of broadcast and unicast services over the same carrier, i.e., while receiving eMBMS broadcast video additional programme information and on-demand content can be requested and served over unicast.

New functionalities have been integrated into the testbed as a result of the 5G-Xcast project. The existing radio network was upgraded several times from its initially 3GPP Release-11 compatibility up to 3GPP Release-14 compatibility. The existing core deployment, which was dedicated lab test software, was replaced by a commercial product, the Nokia Micro Core Network which is a unique integrated – SW only - solution that provides all the Core virtualized network functions (VNFs) in addition with management tool that can be deployed on only one server. The modular architecture allows integration into experimental set-ups.

New functionalities in the context of eMBMS have been provided by Expway with the novelty of integrating MBMS Operation on Demand (MooD) capabilities. With this, it is possible to make the network aware of the concurrent demand by multiple users for a service or content item initially provided with unicast. The detection against a pre-defined threshold can be used to dynamically trigger the establishment of an MBMS user service to offload the unicast network. When concurrent user demand for that same service/content drops below a configurable level, the delivery mode can be configured to transition back from broadcast to unicast.

Expway's software is installed on a virtual machine that comprises the following components:

- The Broadcast Multicast Service Center (BM-SC): the core multicast/broadcast functionalities that receive the content through the xMB interface in unicast and converts into multicast data and sends to the MBMS Gateway (MBMS-GW).

- The MBMS-GW receives the multicast data and forwards it to all relevant eNodeB (eNB) in the network.

- The Broadcast Provisioning Manager (BPM): the provisioning manager that provides the web interface to schedule the eMBMS services and control the BM-SC.

- The Consumption Report and Analytics receive the consumption report messages from the middleware on the phones, analyse the audience size and decide when to switch between unicast and eMBMS delivery.

- The emulated Mobility Management Entity (MME) is used to handle the control plane setup of the eMBMS session. Its purpose is to bypass the MME in case the operator's MME does not support eMBMS.

Expway has provided the Bittium phones with the latest Expway's middleware version that supports MooD functionality. All Expway's software in both terminal and network is compatible with the eMBMS functionalities specified in 3GPP Release-14.

Note that the complete testbed is Release-14 compliant. However, the RAN only supports the configuration of a 15 kHz sub-carrier spacing with extended CP. A diagram of the testbed architecture is shown in Fig. 1.

A further aspect explored in the testbed is the provision of services like those are available today by Hybrid Broadcast Broadband TV (HbbTV) [24]. The live adaptation of a Moving Picture Experts Group (MPEG) Transport Stream (MPEG-TS) to include HbbTV application signalling permits the transmission of regular TV services with the possibility to display data associated to the programme at the UE. This permits to use the smartphone as a TV tuner and provide access to on-demand content without the need of a specific application but a standard one that depending on the service will be able to access different content servers.

In addition, the multi-link capability was also deployed in the IRT Munich testbed to enable the simultaneous use of multiple links to deliver content. This technology is currently implemented over unicast links. End-to-end performance measurements for each link provide different results and cannot be applied to PTM transmissions. The challenge of 5G-Xcast is to extend the multi-link concept to be used for multicast/broadcast within the HBS so that users can benefit not only from seamless transitions between broadcast and multicast to unicast (and vice versa), and/or seamless unicast experience side by side with broadcasts to others, but also high-quality transmissions.

The bonding device or software at the viewing user side communicates with the bonding Gateway (GW) which is on the core network or even remotely, at the publisher or the cloud. These two entities exchange information about the performance of each of the links. The content transmitted from the Gateway down to the viewing device is split over all available links, operators, technologies or Internet Protocol (IP) routes according to their momentary performance. The content is then reassembled at the viewing device as a coherent data stream ready for viewing. The content itself is not manipulated in any way, i.e. the delivery is completely agnostic to the content. Hence complete seamlessness is achieved.

## B. *5GIC Surrey Testbed*

5GIC regards an extensive fibre network as an essential component of 5G networks, where its existing campus-wide 5G testbed based on the University of Surrey in the UK is being further developed to allow 5G technology demonstration of emerging PTM technologies. This aims to support increasing mobile broadband services, speed, capacity and coverage for large numbers of mobile users or devices who are interested in consuming broadcast/multicast content, all by using the same network.

Currently conducting world-wide research and innovation on software defined networks (SDN), network function virtualization (NFV) and mobile edge computing (MEC), the 5GIC testbed's RAN has the flexibility to offer outdoor/indoor services and develops a fully virtualized and componentized network sliced architecture. This supports mainly LTE-A at present and meets the requirements of 5G NR as well as broadcast/multicast specifications. More specifically, the outdoor RAN, covering dense urban, urban, rural and motorways, consists of three macro cells equipped with 8T8R (eight-branch transmit and eight-branch receive) remote radio units/active-antenna-system-enabled RRHs along with 38 small cells with cloud RAN processing unit functions with programmable basebands, while the indoor RAN contains 6 LTE-A small cells and dozens of Wi-Fi access points. For the core network, 5GIC has designed its own fully-featured virtualized evolved package core with 5G Flat Distributed Cloud (FDC) components [25], e.g., cluster member (CM) and cluster controller (CC), as shown in Fig. 2. The 5GIC FDC solution operates as a dynamic "Cluster of Infrastructure" which can be dynamically re-arranged horizontally by topology/ user population load and vertically by network slicing according to user contexts. The architecture involves network entities that can be scaled in and out due to a virtualized implementation, as well as a simple association-based control plane, separated from the user plane which gives faster access, better performance and simple, stakeholder based scalable security. The network is context-aware, thus enabling changing connection points and service and slice provisions dynamically by user context, within radio constraints. The FDC core is also capable of selecting Internet/Intranet breakout points and re-directing traffic according to context and content/applications requested by the users. One of the key innovations is the realization of a control plane node where a non-access stratum (NAS) control plane is integrated with common control signalling (i.e., control plane plus user plane control). This expedited and integrated signalling allows context awareness, and full control plane and user plane separation. This approach improves the access speed and QoE, offering high flexibility and capability to rapidly slice to suit user demographics.

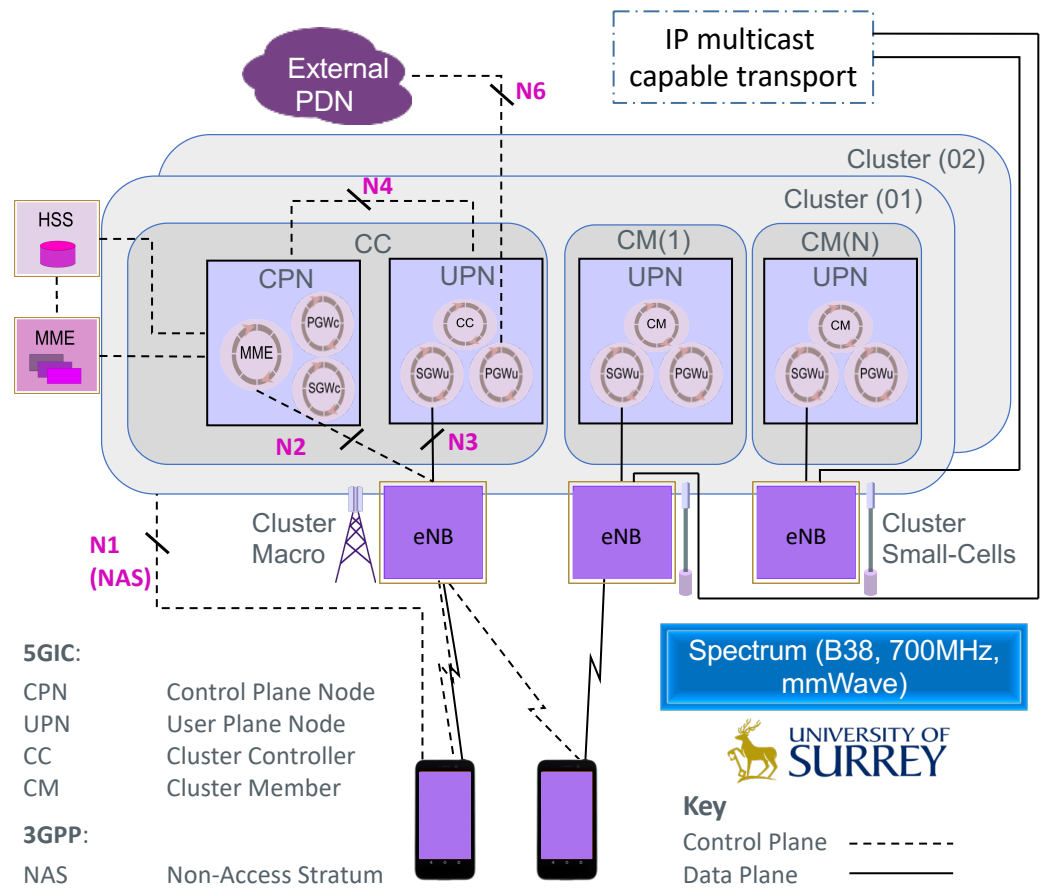

Fig. 2. 5GIC Flat Distributed Cloud (FDC) Architecture with IP multicast functionality.

Leveraging on the enabling functionalities (SDN, NFV and MEC) in the Surrey testbed, 5G-Xcast aims to support new object-based media services which require flexible usage of time-frequency resources and guaranteed latency. For example, in the OBM demonstration, PTM can be used to deliver commonly used and bandwidth-heavy objects to the edge to overcome scarcity of resources in the network. This could either be to a MEC node to overcome bandwidth limitations within the core or directly to the handset to reduce the use of radio resources. Meanwhile, more bespoke, personalized and/or less bandwidth-expensive objects can be delivered over unicast. To this end, IP multicast capable transport is developed in the 5GIC Surrey testbed, as shown in Fig. 2.

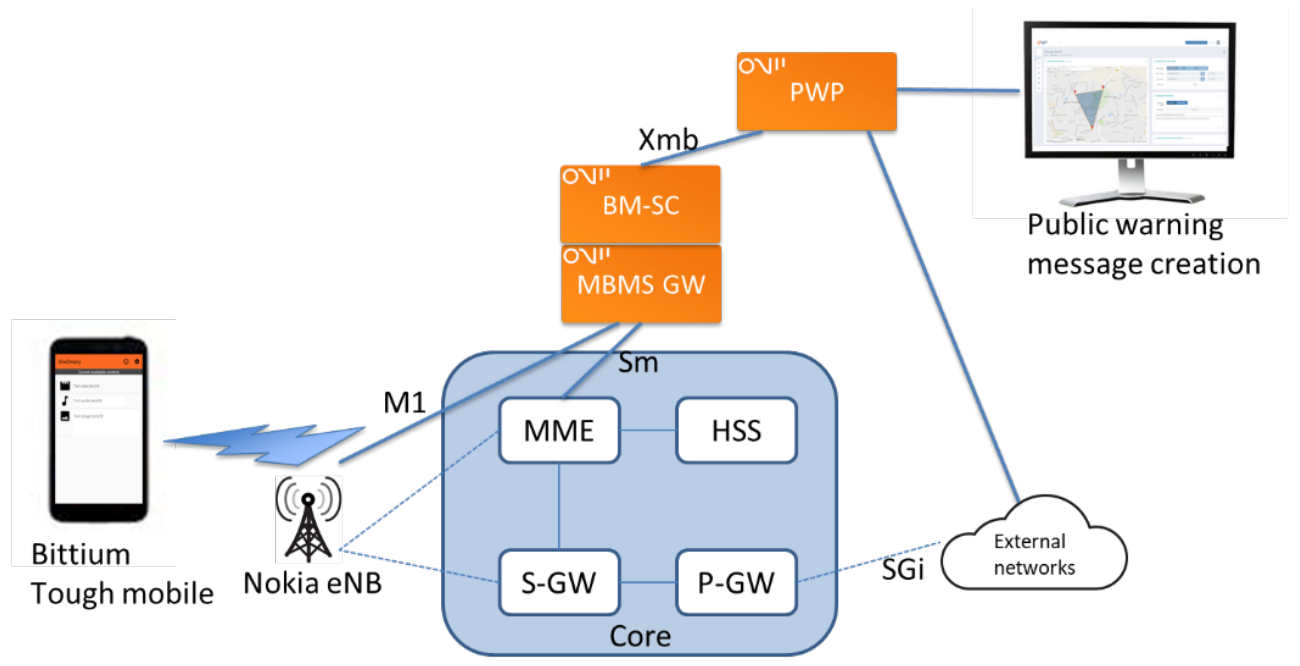

Fig. 3. Architecture for the transmission of public warning alerts in the Turku testbed.

### C. *TUAS Turku Testbed*

The TUAS 5G testbed focuses on spectrum below 6 GHz. Currently the 700 MHz 5G candidate band and the 2.3 GHz Licensed Shared Access (LSA) band are supported for LTE. The 5G candidate band 3.4-3.8 GHz has been added to the testbed. The test network sites are in TUAS campus area in Turku, Finland. The testbed is an integral part of the 5G Test Network Finland (5GTNF) ecosystem [26], which coordinates the integration of the Finnish 5G testbeds. Current 5GTNF testbeds are located in Turku, Espoo, Oulu and Ylivieska.

In addition to the cellular systems, the testbed incorporates technologies such as digital television broadcast network, industrial radio modems, TV White Space radios, LoRa [27]. Further, a spectrum observatory network has been built in the GlobalRF Spectrum Opportunity Assessment project in WIFIUS program, which was jointly funded by the National Science Foundation (NSF) in the US and Tekes in Finland. The project built an international network of RF spectrum observatories continuously collecting long-term spectrum data to study the trends in spectrum utilization and to identify frequency bands where spectrum sharing could be feasible.

The TUAS 5G testbed is developed in the 5G-Xcast project to support eMBMS for demonstrating the PW system. The testbed is updated to contain the necessary network entities, e.g., MBMS-GW and BM-SC. The architecture developed for PW demonstrations is illustrated in Fig. 3.

The alert originator using the one2many Public Warning Platform (PWP) generates the PW messages, which are then transmitted to several users using PTM. The reception of the alert is triggered by a Google Firebase Cloud Messaging (GFCM) push message in the demonstrations using LTE radio. The triggering will be updated for later trials once 5G NR mechanisms are available. This will also allow other types of devices, such as read-only devices with no uplink connection, to receive the alerts. When the alert is triggered, the content is received via eMBMS file download. Finally, the content of the alert message is displayed to the user by a one2many alert application in the receiving device. Commercial Bittium Tough Mobile devices with eMBMS support are used in the PW system demonstration. Further, spectrum manager and multi-link device were integrated in the testbed for demonstrating the delivery of multimedia public warning messages using broadcast, dynamic spectrum use and bonded connections.

### D. *From Testing to Demonstrations*

The development of three testbeds enables testing and trials for establishing the demonstrators and showcase shown in the later sections [28]. The Munich testbed has been utilized as the laboratory for testing HBS with eMBMS, MooD and Multi-Link that brings different demonstrations as presented in Section III, IV, V, VI. The Surrey testbed has integrated OBM functionalities in the 5G system and later this evolves into the Forecaster5G demonstrator as in Section VII. The Turku testbed has served the testing for PW and spectrum management, leading to the demonstrator as in Section VIII.

## III. European Championships 2018: Hybrid Broadcast Service with Linear TV and Add-On Content

The European Championships 2018 provided an opportunity for 5G-Xcast to showcase technical enablers for large-scale real-life media content delivery [29]. A demonstrator on the concept of Hybrid Broadcast Service has been developed by IRT and European Broadcasting Union (EBU), in collaboration with the Nokia and Expway, combining the ability to convey traditional always-on linear TV services in state-of-the-art formats as well as on-demand content, event related information and access to social media.

### A. *Concept and Relation to Use Cases*

The demonstrator presents the following two concepts:

- Live TV content and the signalling for add-on services based on the HbbTV standard are both included in an MPEG-2 TS and transmitted over the LTE eMBMS broadcast system. The broadcast signal is received by stationary eMBMS-enabled TV receivers and by smartphones simultaneously, without the need of unicast connectivity;
- Users can access additional on-demand content either via an HbbTV application on TV sets or a Hyper-Text Markup Language (HTML)-based application on mobile phones. The on-demand content is delivered over the LTE unicast link in mobile networks. This gives an outlook on the coming technology convergence and future 5G capabilities of 5G.

### B. *Technical Description*

The HbbTV concept, which is currently based on Digital Video Broadcasting (DVB) along with an auxiliary broadband connection, is extended to a framework in which the broadcast signal is carried over an LTE eMBMS session and the access to unicast data is provided with LTE unicast. Target receivers under consideration are smartphones and tablets as well as TV-sets.

MPEG-TS is specified as a container format to encapsulate audio and video streams together with programme and system information. Three TV programmes are encapsulated with one of them linked to the HbbTV service information. The main video feed (say linear TV) can be delivered as a standard compliant MPEG-TS. HbbTV services are signalled through the linear broadcast service in the so-called application information table; the application (such as the red button [28]) enables auto-start and is automatically loaded when the user tunes in to the broadcast service.

In this work, two different HbbTV applications are developed, one meant to be displayed on a TV-set and another to be displayed on a smartphone. A demonstration example is shown in Fig. 4 illustrating (from 1 to 8) the different steps and functionalities implemented as follows:

1. The main content is provided by the EBU from the European Championships venues via satellite and is encapsulated in a MPEG-TS alongside other live TV programmes. HbbTV signalling is inserted pointing to additional on-demand content offered by the broadcaster. Note that additional on-demand content is stored on a dedicated web-server.
2. A small-scale computer-based solution including LTE Evolved Packet Core (EPC), MBMS and radio stack permits the delivery of the broadcast signal (MPEG-2 TS over Real-Time Transport Protocol - RTP) over LTE downlink and the allocation of the remaining unicast capacity for on-demand traffic.
3. Internet connectivity provides access to the servers with on-demand content. Broadcasters can direct users to their own content repositories.
4. Smartphones with Expway's middleware allow users to watch live TV programmes via the broadcast system and on-demand content via a mobile web application and a unicast internet connection.
5. A smartphone acting as a set-top-box forwards the original MPEG-2 TS to a TV-set that can tune in to the live TV signal with the possibility to access HbbTV services
6. The EPC is connected to the eNB that serves 4 RRHs in the urban and rural areas of the city.
7. The signal is transported to the RRH which can transmit the radio signal.
8. On air, an LTE sub-frame is configured with 60% capacity dedicated for broadcast and 40% capacity for unicast traffic.

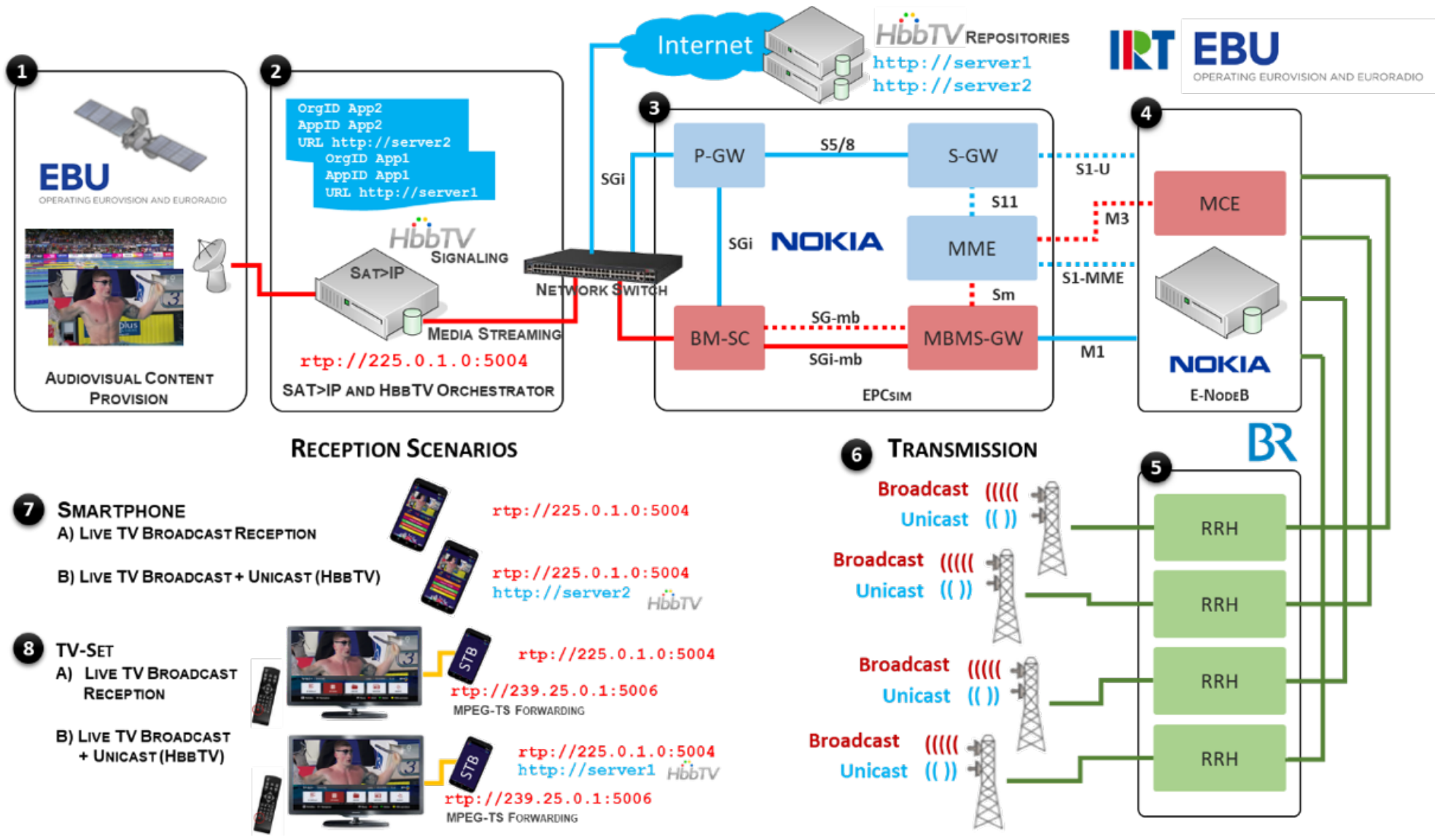

Fig. 4. HBS showcase setup for European Championships 2018.

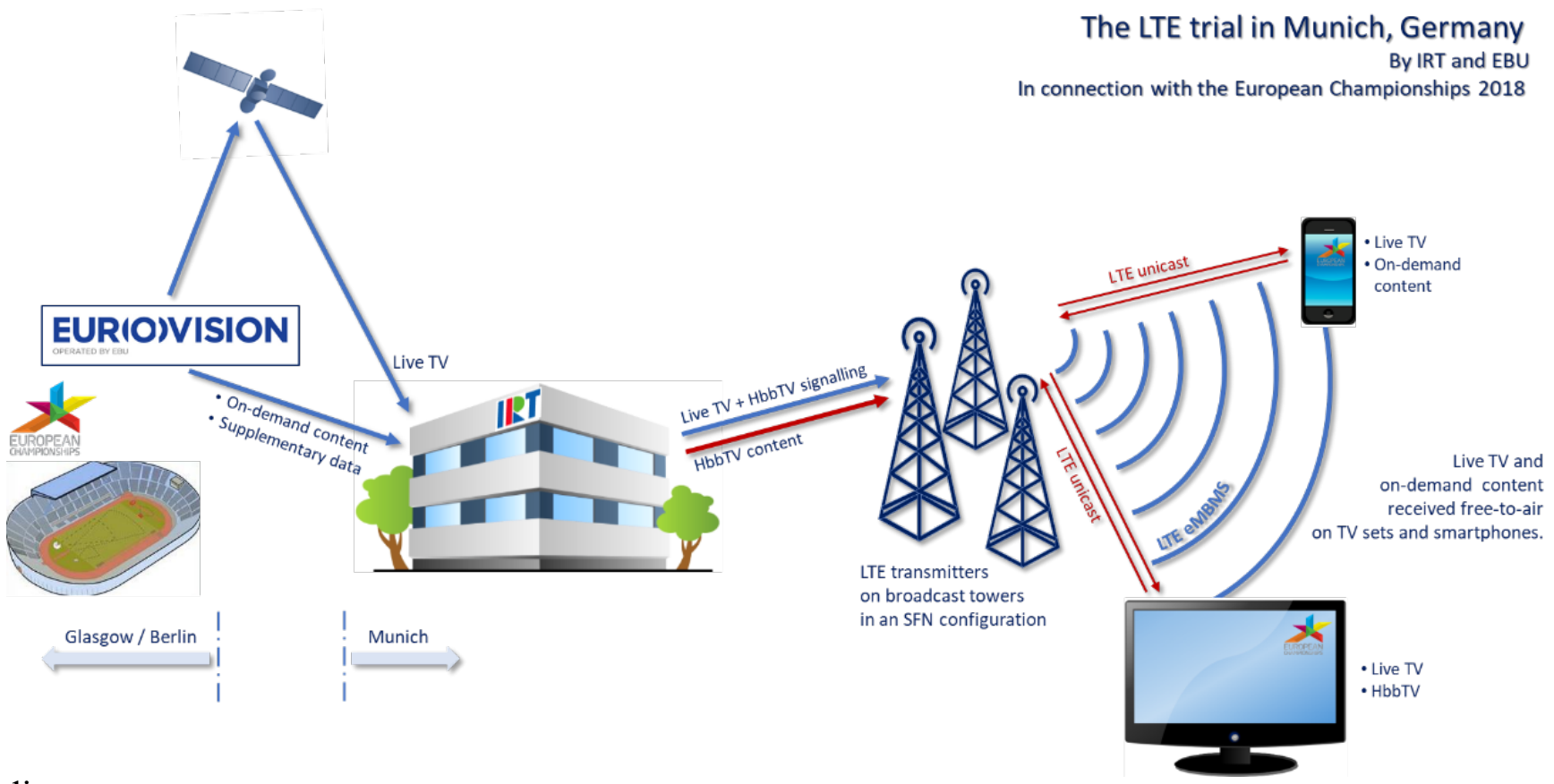

Fig. 5. HBS showcase block diagram.

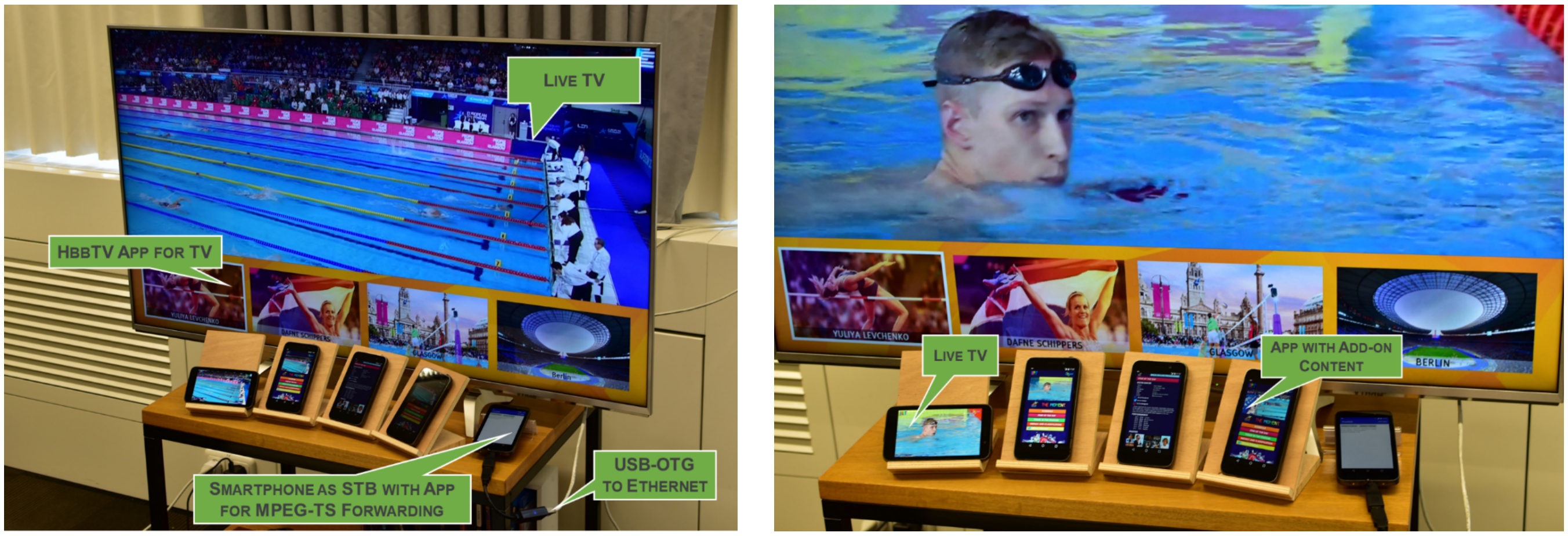

Fig. 6. A demonstration example of HBS showcase at the European Championships 2018.

With this setup it is possible to highlight operation modes available in LTE eMBMS Release-14. By using dedicated broadcast carriers as introduced in LTE eMBMS Release-14, with self-contained signalling providing no need to register the UE, of uplink, or of a SIM card. The additional unicast connection can be provided by means of a regular unicast carrier (e.g. LTE unicast). From the Core Network point of view, this showcase also demonstrates the possibility of operating a dedicated broadcast network only with limited functionalities (e.g. MBMS-GW, BM-SC), without the need of deploying the full set of functionalities from the core network when unicast traffic is not going to be provided. Fig. 5 shows an overview of the showcase using the Munich testbed.

### C. *Demonstrations and Results*

One of the key points of this showcase is the development of two applications for the smartphone for the demonstration purpose. One application is developed in order to provide hybrid services (the combination of linear and non-linear content) directly on the smartphone so that different TV services (within the MPEG-TS) over eMBMS can be tuned and displayed on the screen. The first app is able to identify when HbbTV signalling is present in the stream so that access to additional content over the internet can be provided. The add-on content showed in the smartphone is retrieved from a web server where an HTML site has been developed by the service provider allocating the desired information, in this case, related to the European Championships (statistics, information about sports, schedules, etc). The other app is developed in order to forward the MPEG-TS to an IPTV-set able to show the live TV programme and also an HbbTV app prepared on purpose with the correct format to be displayed on the TV-set.

A demonstration example is presented in Fig. 6. Given that the 5G technologies were not fully developed at the time of the European Championships 2018, the showcase incorporated the available pre-5G technologies. However, the showcase still enables us to explore the ability of current and future 3GPP based 4G/5G specifications to provide an HBS user experience in a fully wireless environment combining broadcast and unicast transmission to, first, access the broadcast signal and, later, establish a unicast connection to access on-demand and added value content from the HBS-server.

## IV. Demonstrator: Large Scale Media Delivery powered by MooD and Free-to-Air Distribution to TVs and Smartphones

Here the primary objective is to demonstrate the capability of 5G technology to deliver traditional always-on linear TV with added-value content on-demand and, furthermore, the scalability of the 5G system providing sustained quality of experience for increasing demands of high-quality content and the seamless switching between broadcast and unicast delivery modes powered by MBMS operation on Demand.

### A. *Concept and Relation to Use Cases*

For a large-scale distribution of popular content such as live sports, media delivery is not always based on broadcast technology, but unicast transmissions can also play a role in the complete system in order to serve specific demands from individual users. The demonstrator aims at showing: 1) when unicast and broadcast capabilities are essential for an efficient provision of media content; 2) how the eMBMS technology (i.e., LTE Broadcast) can optimize the network resources for the mobile network operator during the important events with the aid of MooD.

### B. *Technical Description*

5G technology should be able to support the delivery of live TV programmes to user devices with 1) a broadcast mode where the live TV programme is delivered according to Quality of Service (QoS) and coverage requirements defined by the service provider, 2) an adaptive unicast/broadcast switching mode where live TV, encoded with multiple Dynamic Adaptive Streaming over HTTP (DASH) profiles, can adapt to user demand and reception conditions.

#### 1) *Adaptive Unicast/Broadcast switching according to demand*

A live stream is originally delivered over unicast to a few smartphones. The content then becomes popular, the MooD feature is activated, and the network automatically switches to broadcast mode to optimize the overall system resource usage and to guarantee the QoS. The smartphones with eMBMS middleware installed automatically switch to broadcast if they are in the coverage area of that signal. The switching is transparent to the users who do not experience any interruption while watching the content. When the content becomes less popular, the network automatically switches back to unicast while ensuring the smooth playback experienced by the end users.

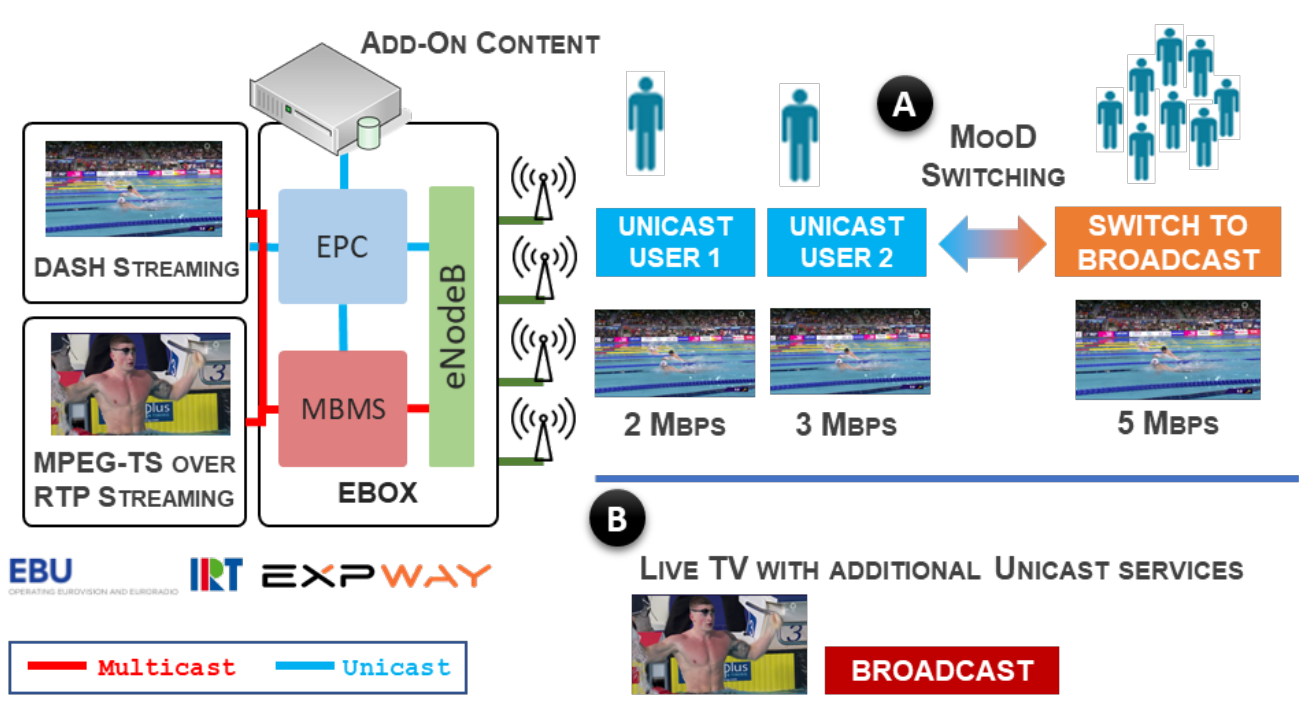

Fig. 7. An ecosystem for demonstrating a large-scale media delivery powered by MooD and free-to-air distribution to TVs and smartphones.

*2) Live TV Broadcast with embedded HbbTV signalling*

Recall the first HBS demonstrator as shown in Section III; in this demonstrator live TV content and the signalling for the access to add-on services based on the HbbTV standard are both included in an MPEG-2 TS and transmitted over eMBMS. The broadcast signal is received by stationary eMBMS-enabled TV receivers and by smartphones simultaneously and without the need for unicast connectivity. Users can access additional on-demand content either via a HbbTV application on TV sets or an HTML-based application on mobile phones. The on-demand content is delivered over the unicast link in the mobile network.

Two different types of content should be made available: the main live TV programme to be broadcast and a stream of a second programme, encoded with DASH with multiple profiles to adapt user requirements. Both services streamed via unicast and live broadcast TV can be delivered in parallel using a maximum of 60% (up to 3GPP Release-14) or 80% (from 3GPP Release-14) system resources for broadcast within the same system.

### *C. Demonstrations and Results*

Similar to the HBS demonstrator and showcase as presented in Sections III, this demonstrator makes use of the original content provided by the EBU from the European Championships 2018 encapsulated in a MPEG-TS, displayed on IRT HbbTV apps. Expway also provides an all-in-one system called eBox, which enables broadcast capabilities powered by the Expway's BM-SC in the core network and the Expway's middleware on the smartphones. An ecosystem is illustrated in Fig. 7.

The demonstration is aiming to employ the commercial equipment enabled with the specific middleware to realize the operations explained above. Smartphones with Expway's middleware allow users to watch live TV programmes via the broadcast system and on-demand content via a mobile web application and a unicast internet connection. Users can select to watch the live TV programme via broadcast or chose another stream via a unicast link. When the audience consuming the unicast stream increases over a certain threshold, an MBMS session is established thanks to MooD feature so that the users transparently switch between unicast and MBMS. When the demand decreases, users can again be served via unicast. The reception of the live TV programme over broadcast is not affected by the increasing number of users.

## V. Demonstrator: Converged Autonomous MooD in Fixed/Mobile Networks

British Telecom (BT) and Expway have worked together to produce this demonstrator, to show enabling features proposed by 5G-Xcast from the content distribution framework (CDF) point of view [30]. We highlight the following features to be demonstrated through the integration of the live, unmodified, BT Sport commercial service with the 5G-Xcast CDF:

- The use of multicast/broadcast as an internal network optimization, rather than as a service to be sold.
- The use of simple unicast interfaces with content service providers to simplify integration and facilitate adoption.
- How client applications do not require any modification to benefit from the use of multicast/broadcast.
- How the framework is applicable to both fixed and mobile networks.

### *A. Concept and Relation to Use Cases*

The concept of this demonstrator is to realize implementations of the CDF on both fixed and mobile networks. We aim to gain insight into the practical challenges of implementing the framework and testing the feasibility of various approaches. The resulting implementations potentially allow the benefits of the framework to be communicated in a visual and engaging way.

We choose to build two instances, one for the fixed network and one for the mobile. While they share similar components, they have different requirements in terms of network architecture. For example, the fixed network features a home gateway which terminates a broadband connection and allows in-home devices to connect wirelessly over Wi-Fi, while the mobile network includes a cellular base station that provides a radio connection direct to the mobile UE.

The demonstrator can show how the framework is able to steer unicast stream onto multicast/broadcast delivery as a function of the audience size. This can be accomplished by using MooD and allows video playback to continue uninterrupted during a delivery mechanism switching event.

### *B. Technical Description*

The configuration for the mobile demonstrator is shown in Fig. 8. It comprises a content source in HTTP live streaming (HLS) format which the BM-SC ingests in response to configuration information sent to it by the broadcast provisioning manager. The BT Sport App is installed on an MBMS capable device which also includes an adaptation layer and Expway's MBMS middleware (shown as Multicast channel Scheduling Period-MSP Server in the diagram). When the BT Sport App makes a request for content, this is detected by the request routing component that redirects the request to the adaptation layer. Here, the request is inspected and the appropriate MBMS service is activated through the use of the MSP client. This causes the HLS content to start being received and is placed in the MSP Server cache where the adaptation layer is able to request it and forward it on to the BT Sport App. The app remains unmodified and is not aware of the use of multicast/broadcast. The adaptation layer allows it to operate as a simple unicast client.

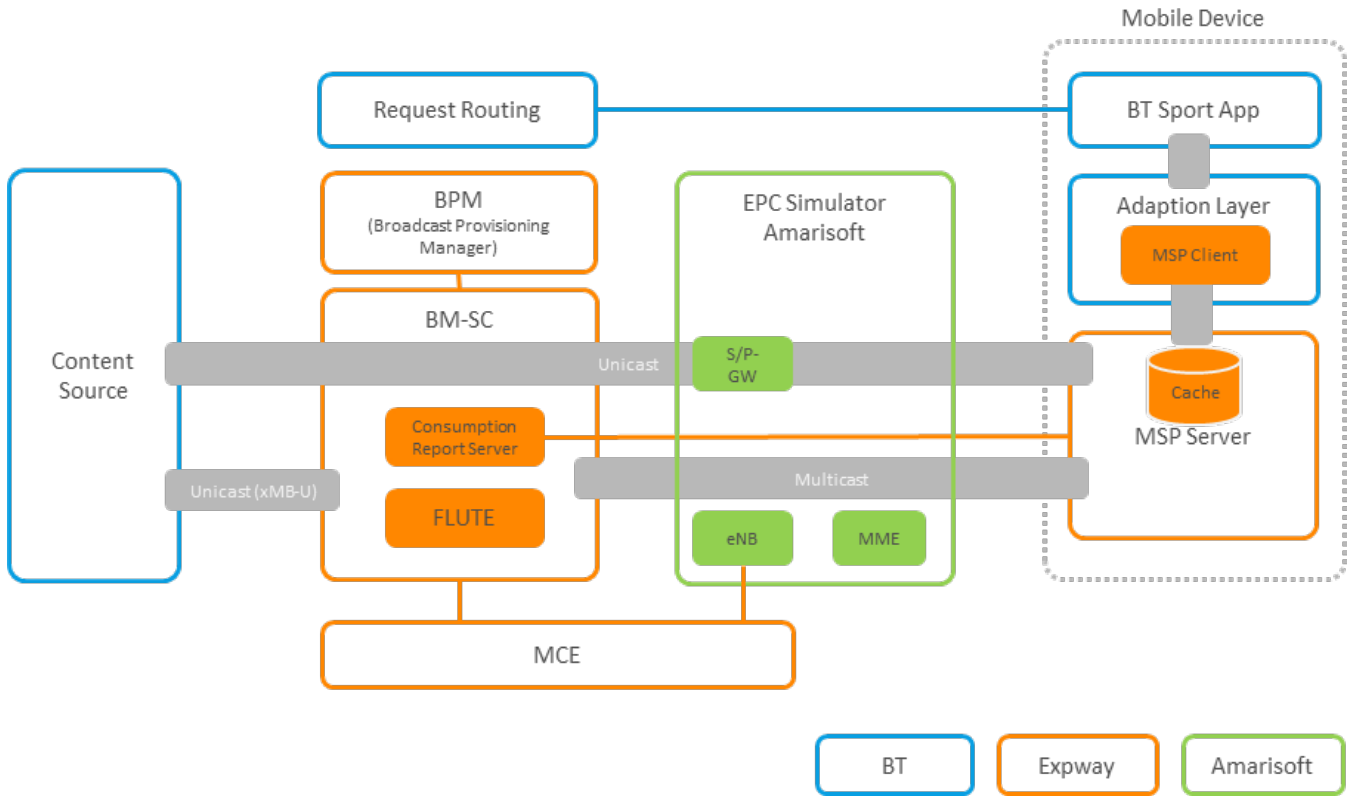

Fig. 8. Demonstrator configuration for mobile case.

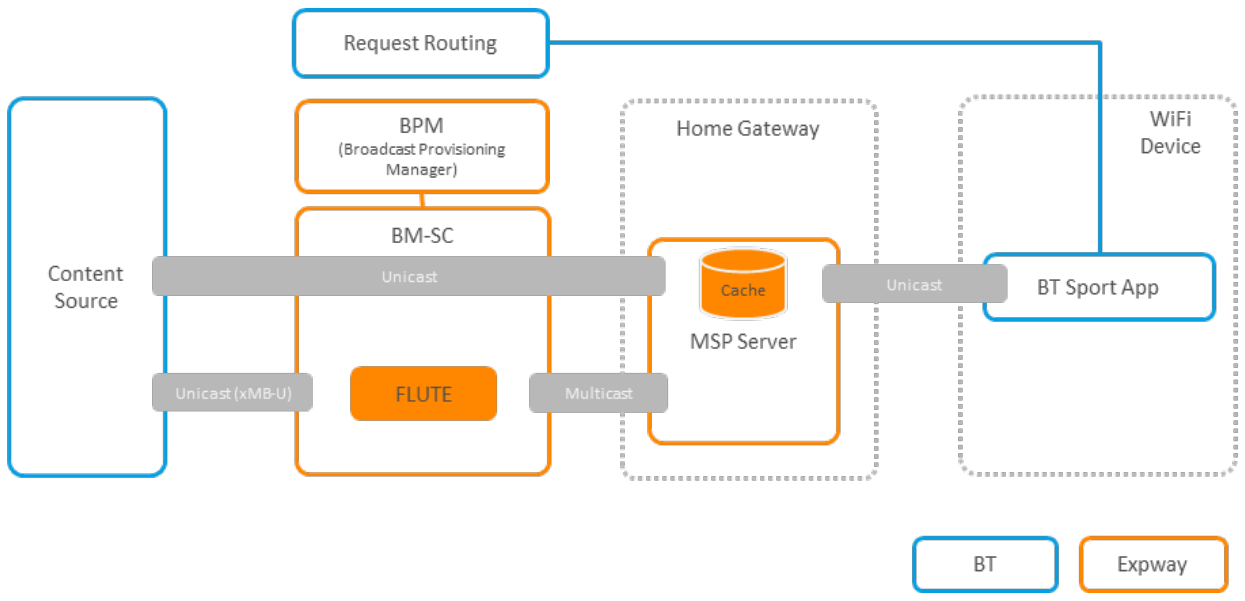

Fig. 9. Demonstrator configuration for fixed network.

MooD is implemented in the BM-SC where a consumption reporting component monitors the demand for content and through the multicast coordination entity (MCE) is able to activate and deactivate MBMS services. Thus, multicast/broadcast is used as an internal network optimization leaving the content service provider and client application with simple unicast interfaces. The mobile network aspects of the demonstrator are implemented using Amarisoft EPC components.

The approach taken for the fixed network demonstrator is largely similar, although slightly simpler as shown in Fig. 9. The same content source and BM-SC/BPM components are present and are configured to ingest the HLS content and stream it out on a multicast address as configured through the BPM. In the fixed network, the introduction of a home gateway (HGW) affects how content is received. When the app requests content, the request routing component directs the request towards the MSP server running on the HGW. An adaptation layer is not required as the MSP server is able to match the requested URL directly to an appropriate multicast address. When the MSP server makes the match, it joins the appropriate multicast address and starts to receive media segments which are placed in the cache. The segments are subsequently forwarded to the app as HTTP responses. Initial requests made to the MSP server may need to be satisfied by unicast prior to the cache being populated with the required content.

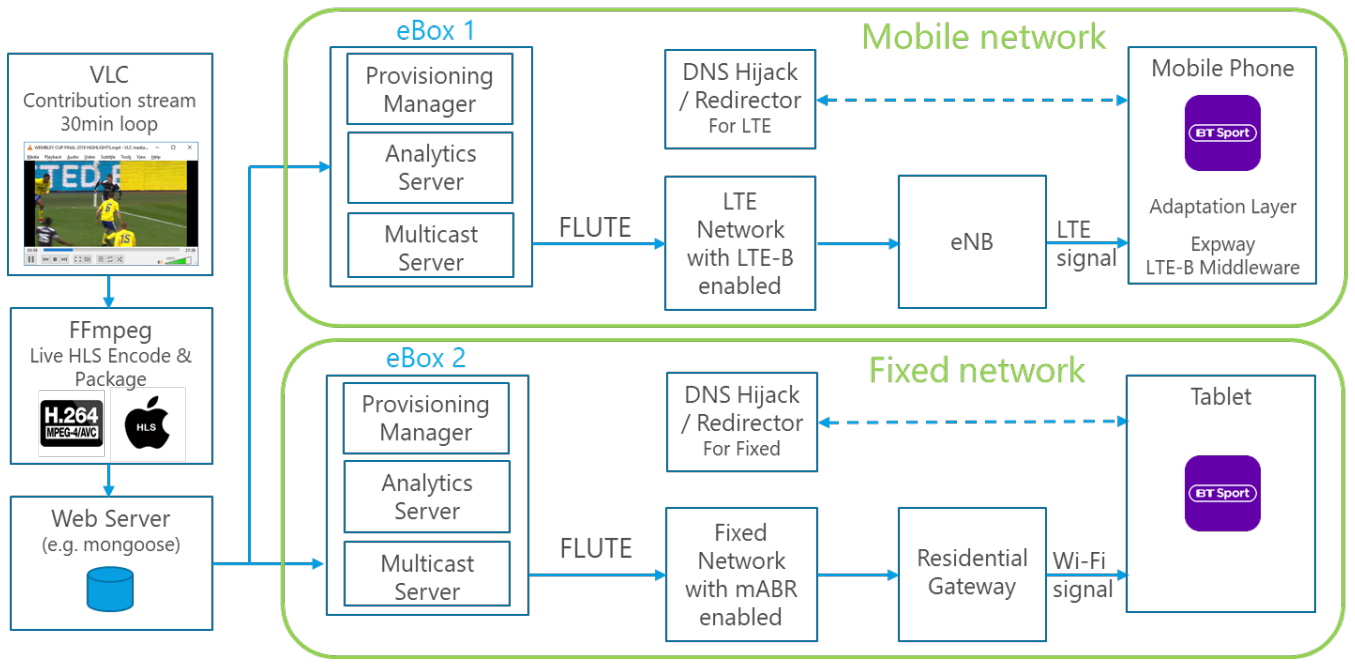

Fig. 10. Demonstrator architecture.

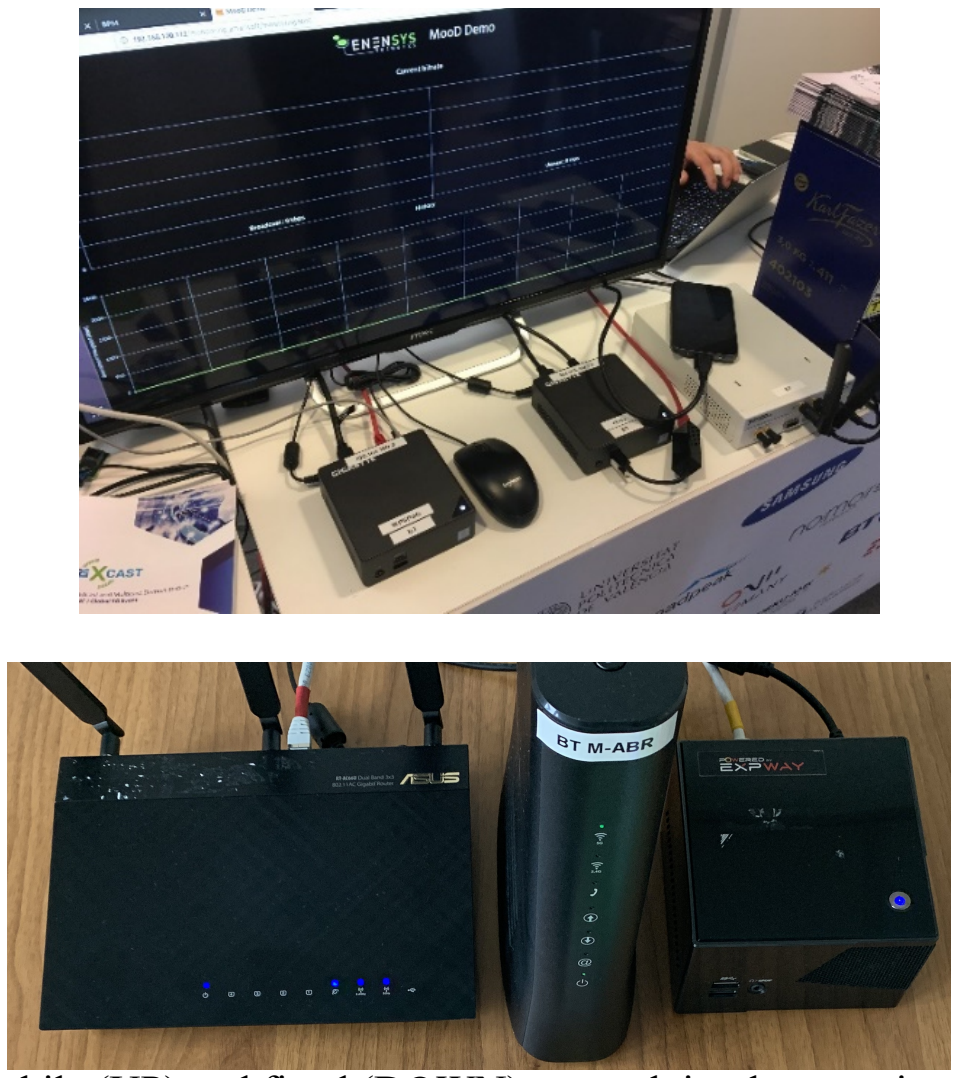

Fig. 11. Mobile (UP) and fixed (DOWN) network implementations.

### C. *Demonstration Setup and Results*

For the demonstration, a high-level architecture of the demonstrator is shown in Fig. 10. Local content is provided using VLC to playback a 30 min football sequence on a loop. This is encoded and packaged by the FFMPEG open-source multimedia suite, and made available to the framework through a webserver. The content is generated locally to avoid any issues around content rights, particularly as the original content streams are intended for the UK while the demonstrator may be shown elsewhere. The eBox components in both implementations ingest the unicast content and prepare it for delivery over multicast. It also includes an analytics server for monitoring the audience size. The provision manager provides a means for configuring and scheduling streams. In the mobile case, an LTE network is provided using Amarisoft software coupled with a software defined radio unit programmed as an eNB. In the fixed case, the network is terminated with a Wi-Fi-enabled residential gateway, as presented in Fig. 11.

The BT Sport application is installed on Bittium mobile devices in the mobile case and tablets in the fixed case. The Bittium mobile devices are MBMS-enabled out of the box. In both cases, real-time visual indication of the load on the network is provided, as in Fig. 12. This shows how unicast delivery can be used when one device is consuming content, but that broadcast delivery is used when two devices are consuming the same content. The example results in Fig. 12 apply for both fixed and mobile networks, showing how the audience size can be used to make the decision about when to switch content onto multicast/broadcast. More details and visual verification for this demonstrator can be found online [31], which also verifies that the switch between delivery modes does not impact on the continuous video playout experienced by the end user.

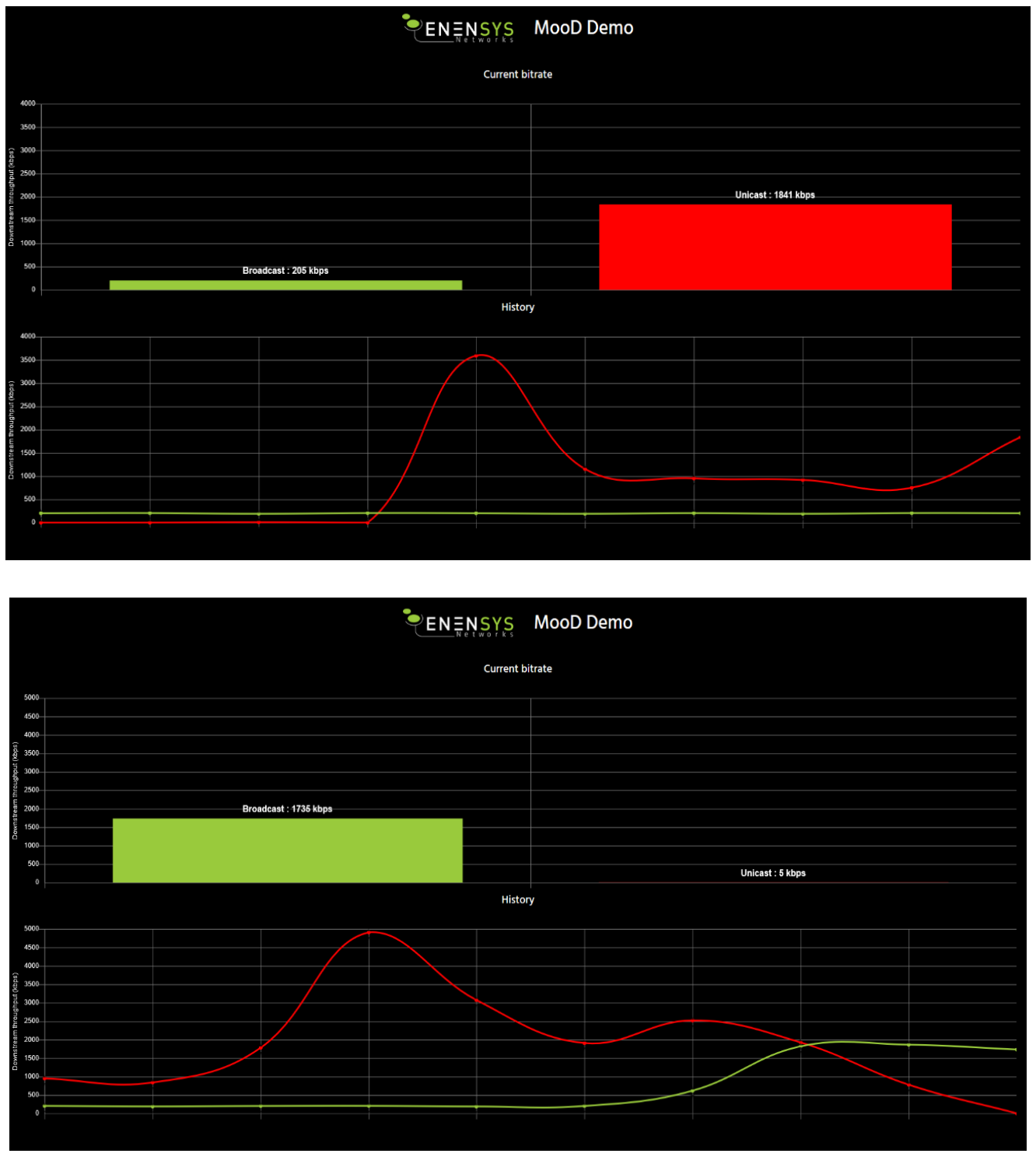

Fig. 12. Unicast delivery of a single live stream (UP) and broadcast delivery to multiple devices (DOWN).

## VI. Demonstrator: Hybrid Broadcast Service with Multi-Link

The demonstrator is developed by Bundleslab and IRT with the support of LiveU and the EBU. The main objectives of the demonstrators are to:

- Improve reliability, bandwidth, mobility and traffic optimization by multi-link connectivity between different radio access technologies and networks.
- Create a virtual single broadband connection by the simultaneous use of multiple networks in a dynamic way.
- Improve user experience when moving from outdoor 4G/5G connectivity to indoor Wi-Fi with a seamless viewing experience.
- Demonstrate Multi-Link protocol for enhanced broadcast delivery using on-demand video stream repair via unicast.

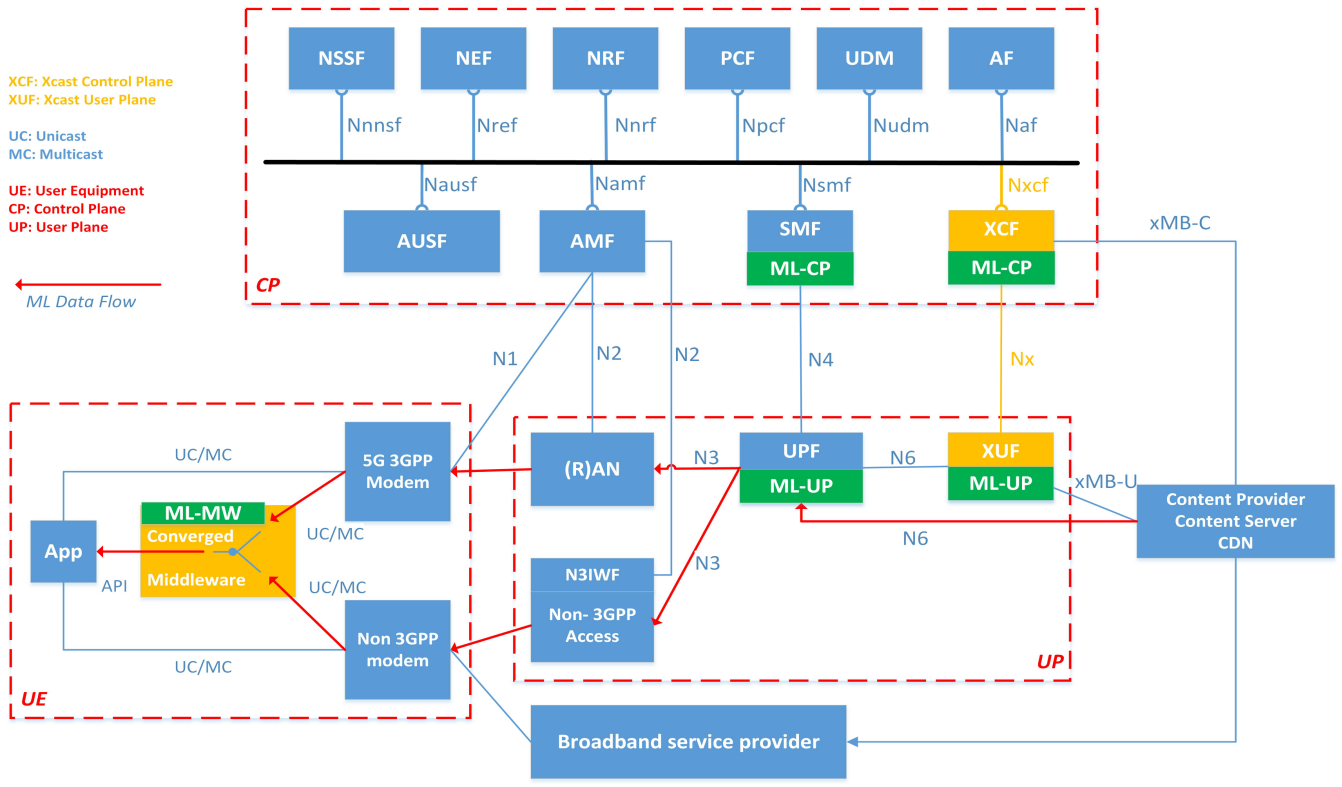

Fig. 13. A block diagram for the demonstrator of Hybrid Broadcast Service with Multi-Link.

### A. *Concept and Relation to Use Cases*

The overall concept of the demonstrator can be explained in the following steps: a UE such as mobile device is on the edge of the broadcast or multicast area experiencing poor broadcast or multicast service or a mobile user is going in and out of that broadcast or multicast coverage area; the content transmitted from the Multi-Link Gateway (ML-GW, the server) down to the viewing device is split or duplicated over available links being them networks from different operators or even using different technologies. The ML-GW is able to dynamic balance traffic according to reception performance. The decision whether to split or to duplicate depends on the desirable gains in throughput, ancillary information and reliability, and a function of the link conditions; the content is reassembled at the viewing device (with eventual duplicates removed) as a coherent data stream ready for viewing. The content itself is not manipulated which means that the delivery is completely transparent to the content.

By means of multi-link it is possible to show the possibility to achieve:

- Better reliability and availability of the service against fluctuation in bandwidth, latency or error rate and enabling a seamless transition between single-Layer 2-link and multi-link could be achieved in a reliable way due to the use of simultaneous multiple networks.
- Increased bandwidth with the possibility to deliver broadband content that would be impossible to deliver over a single link.
- Better mobility support with seamless transitions between coverage areas of different networks or technologies, with continuous QoS and QoE.

### B. *Technical Description*

The architecture of the demonstrator is based on the concepts developed in 5G-Xcast as shown in Fig. 13, where the key blocks are explained below:

- ML-CP: additional functionality in the control plane of the mobile core network, which performs the estimation of QoS parameters (e.g., latency, jitter or datagram loss) for data transfer via each available link, multi-link session setup and release;
- ML-UP: additional functionality in the user plane of the

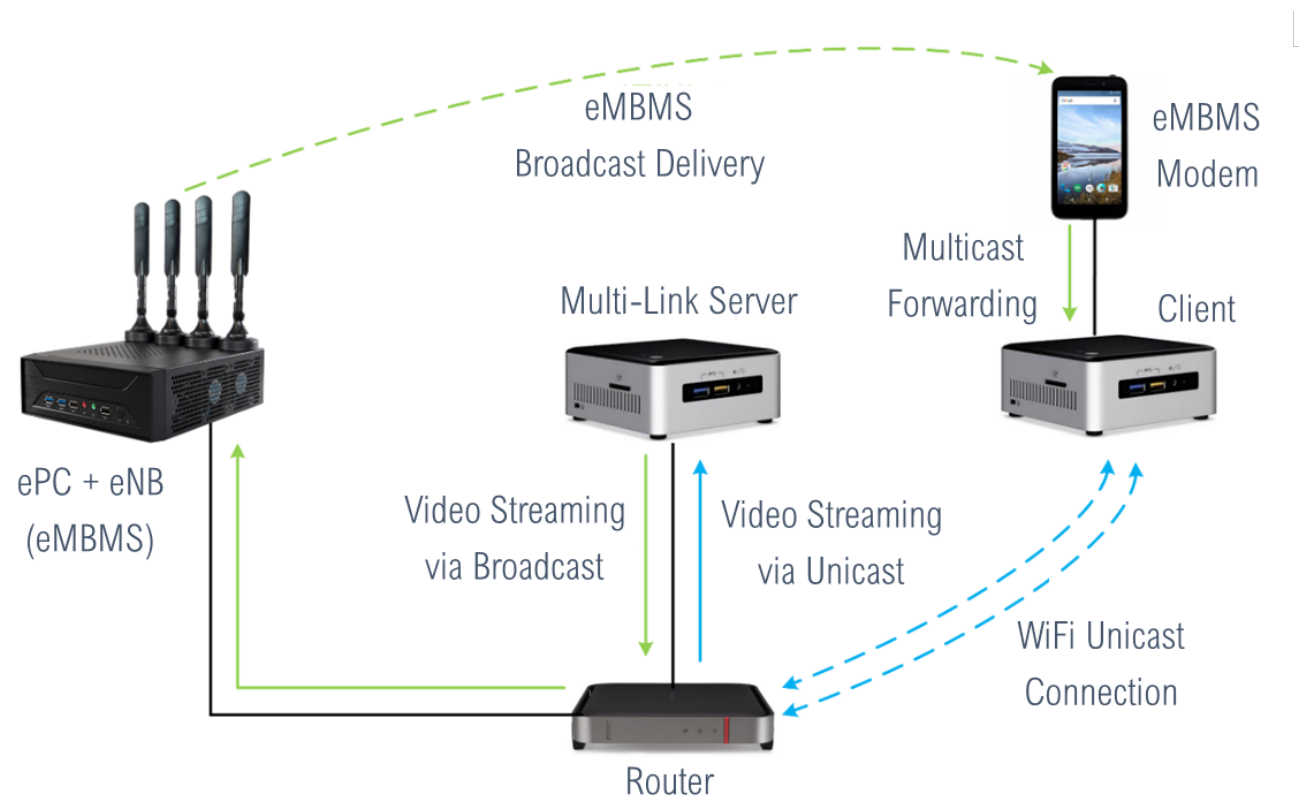

Fig. 14. Demonstrator setup and hardware components of Hybrid Broadcast Service with Multi-Link.

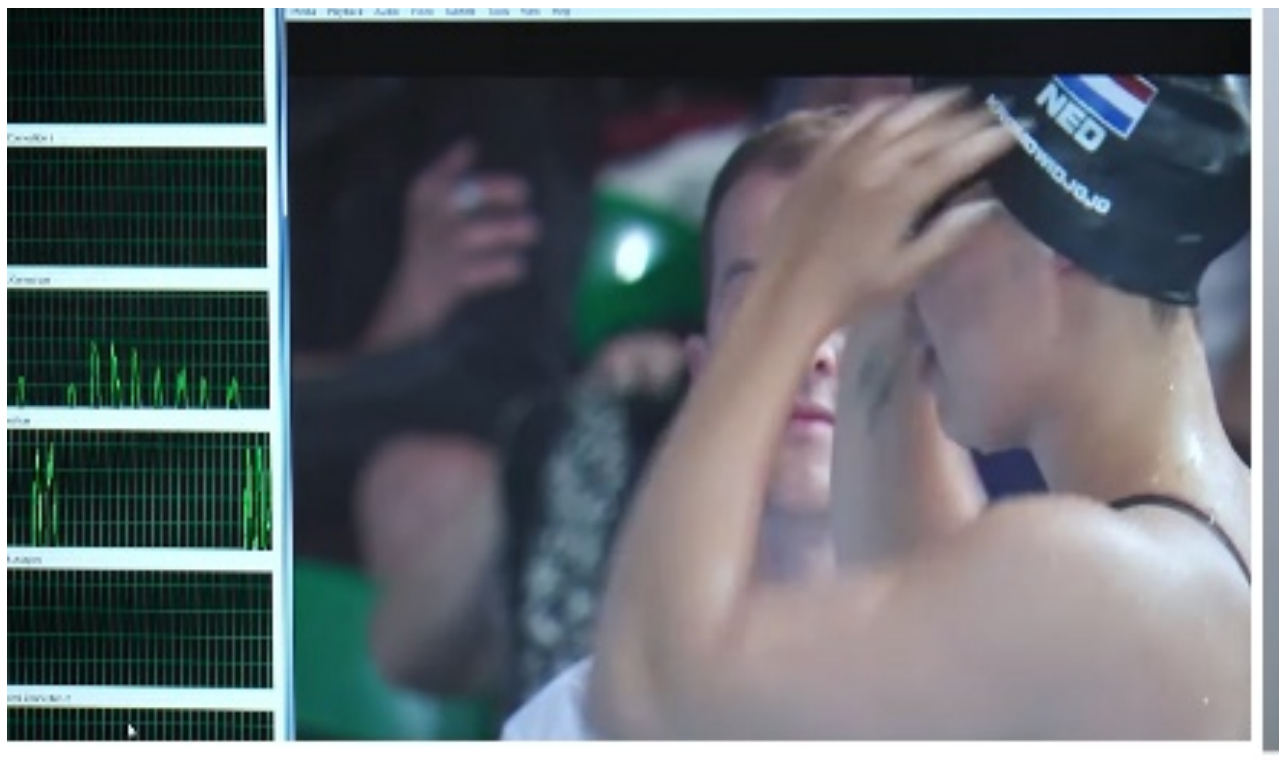

Fig. 15. An example demonstration of BS with Multi-Link.

mobile core network, which performs data splitting, IP tunnel establishment;

- ML-MW: ML middleware (ML-MW) functionality in the UE between the Application and the lower transport levels, which performs data combining, signalling (channel quality data transmitting), caching, providing ML session setup request (QoS parameters).
- ML-GW is able to reroute the data packets through the different available links, and the ML-MW performs the adequate data merger operation at the UE. The ML-MW at the viewing user side communicates with the ML-GW which can be located either at the core network, the publisher, or the cloud. These two entities (ML-GW and ML-MW) exchange information about the performance of each link.

### C. *Demonstration Setup and Results*

The demonstrator setup is shown in Fig. 14, where the hardware components are depicted. The multi-link server is able to reroute the data packets through the different available links, and in this case, Wi-Fi as the unicast connection and a multicast session over eMBMS. At the user end, the multi-link client performs an adequate data merger operation between the traffic delivered via multicast and unicast. In our case, the client is able to analyse the multicast traffic from the eMBMS connection and evaluate packet losses. In the case that a packet drops and errors occur in the eMBMS multicast stream, a repair mechanism starts and the missing packets/chunks are requested to the server and sent via the unicast link. An example result shown in Fig. 15 indicates improved reliability by multi-link bonding at the user device.

## VII. Demonstrator: "Forecaster5G" Object-based Media over Multicast and Unicast

British Broadcasting Corporation (BBC), with support of 5GIC, has developed a demonstrator to show an OBM approach to content delivery. This demonstrator has two objectives:

- Delivering an enhanced audio/video media experience, in which the presentation of the content adapts to the user's environment, the user's preferences, the device's capabilities and includes personalization. This is achieved by using an object-based approach, in which the media is split into multiple objects (e.g. video, audio, images, subtitles, etc.). These are then independently delivered to the device, where they are rendered adaptively.
- Reducing the resource cost of delivering high quality personalised live media content to a large audience over IP by using multicast for bandwidth-heavy and/or commonly used objects, as opposed to conventional unicast. This is achieved reliably by using the Dynamic Adaptive Streaming over IP Multicast (DASM) system, developed by the BBC's R&D department [32].

The demonstrator consists of an object-based weather forecast, thus named "Forecaster5G", in which we have split a traditional weather forecast into multiple objects. In this case the objects include an MPEG-DASH video of the presenter, the presenter's audio, an MPEG-DASH video of a sign language presenter, subtitles, and image assets (which include some personalization). We deliver the MPEG-DASH videos over multicast, via the DASM system, while delivering the image assets over unicast. The objects are seamlessly combined on the UEs into a single, responsive, experience. The OBM using DASM system has been trialled in the Surrey testbed, where we are able to integrate the DASM system into the 5G network within hours, thanks to the network features SDN/NFV/MEC. Here, the architecture (see Fig. 16) and setup are close to that described in Section IV of [33], however, in this instance, the Generic Routing Encapsulation (GRE) tunnel terminates on a test server at the exhibition venue. The multicast packets within this tunnel are then decapsulated and passed to the DASM

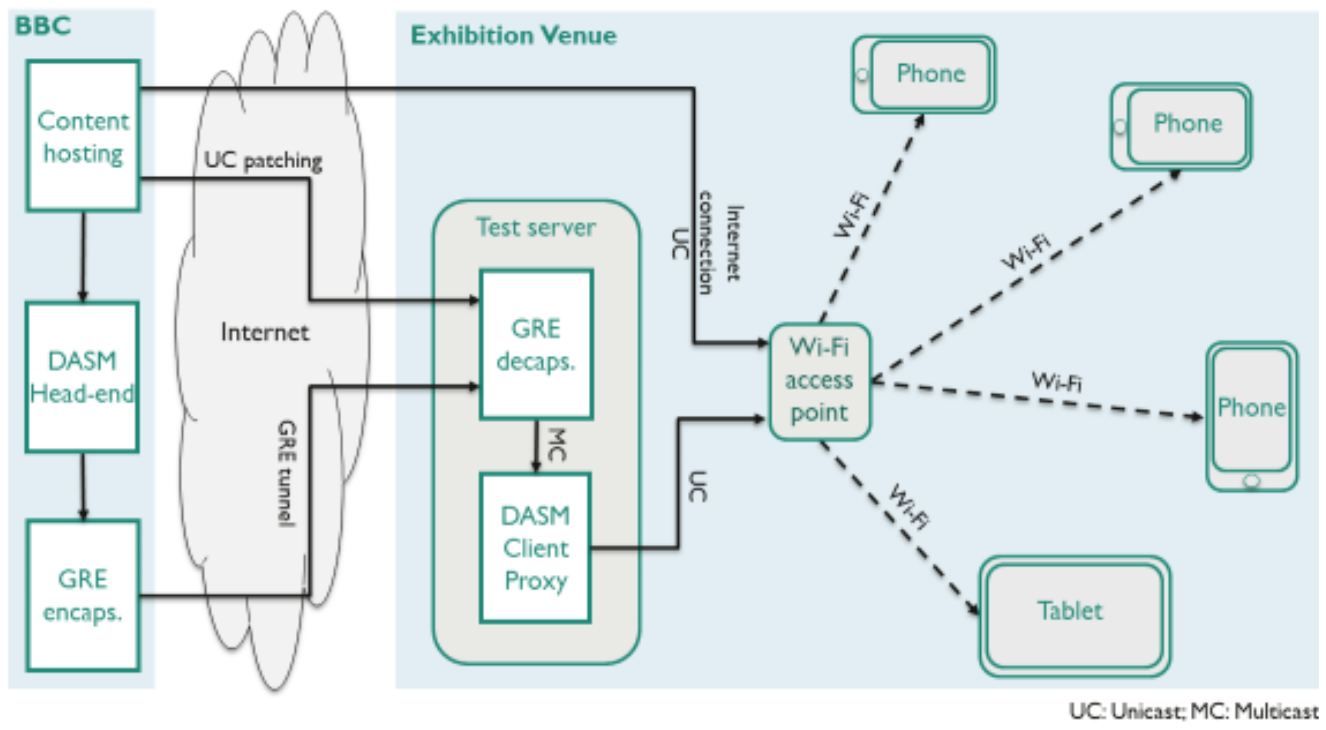

Fig. 16. Architecture of the OBM demonstrator.

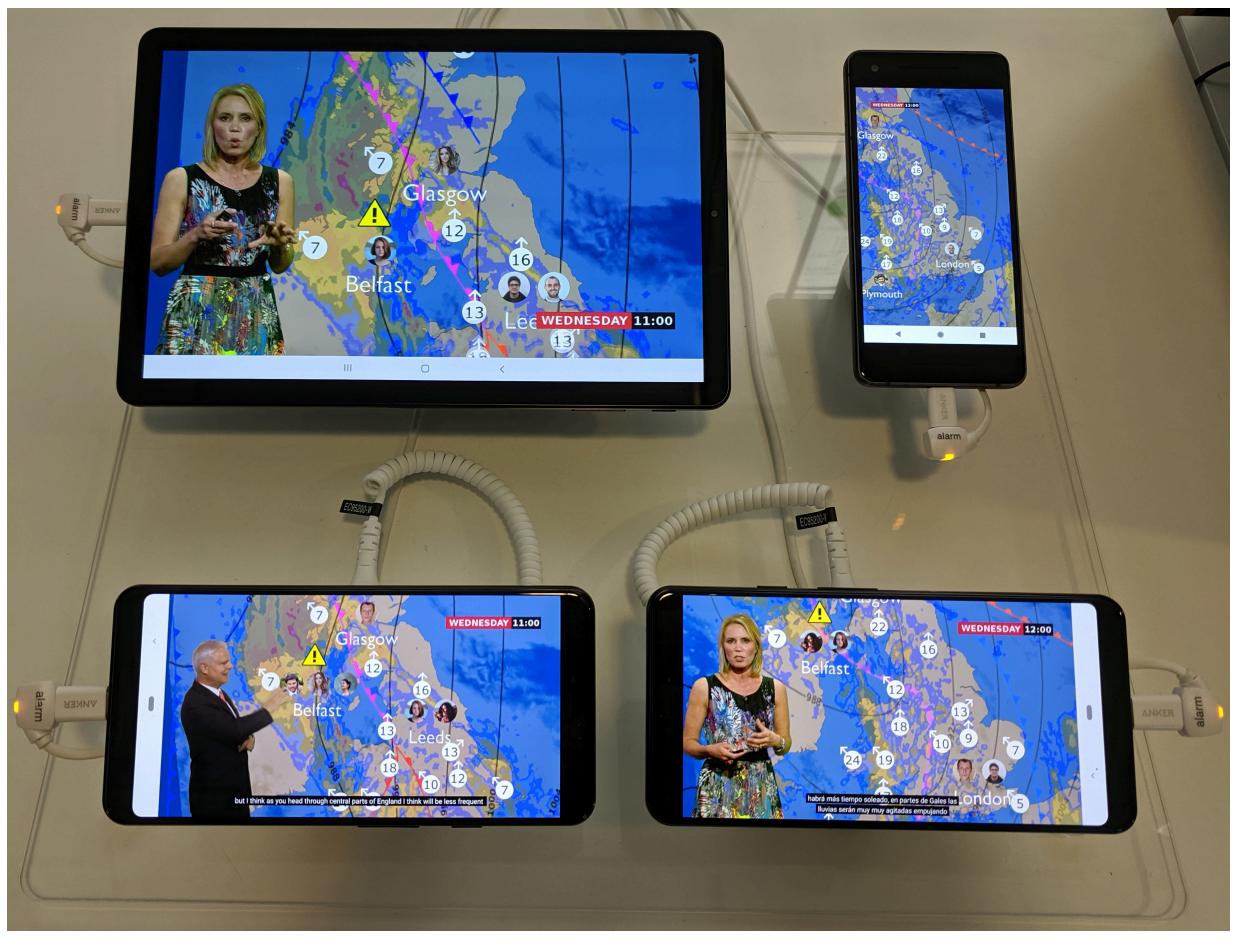

Fig. 17. OBM demonstrator "Forecaster5G".

Client Proxy. As indicated in Fig. 17, the UE then renders the objects depending upon several factors, including user preferences (for example choosing between the main presenter or a sign language presenter), device orientation (portrait or landscape), and unique personalization to the user (in the form of icons for the location of the user's friends on the weather map) [20][21].

## VIII. PW Demonstrator: Multimedia Public Warning, Spectrum Management and Multi-link

For the multimedia PW demonstrator, the main objectives are to showcase various capabilities to deliver multimedia public warning efficiently and know how to include relevant information in the warning for people who are, for example, hearing-impaired or visually-impaired, as well as including additional context information.

To allow reliable transmission of the multimedia public warning alert, transmission of the alert combining broadcasting, dynamic spectrum use and bonding several connections using multi-link is demonstrated. With broadcast, the maximum number of users can be reached while avoiding congestion in the network. Not all users necessarily have broadcast reception capable terminals, and therefore the alert is fetched over unicast connection by such terminals. Further, to increase the capacity, spectrum management is utilized to allow for dynamic addition of capacity bonded in the demonstration with multi-link.

The public warning application supporting multimedia content in the UE is developed, to provide illustrative information for the alert recipient [21]. Information such as maps/pictures/audio/text was demonstrated. The demonstration system includes a default network (Public LTE) and an additional network on Band 28 to demonstrate the performance improvement of ML and dynamic spectrum management, as shown in Fig. 18. In this system, a new spectrum resource is defined and created. The additional spectrum capacity is added to the default LTE link as a second link in the same network by using the bonded multi-link. The trigger for the PW alert is sent using the Public LTE to the UE. The UE receives the multimedia components primarily using the eMBMS broadcast. Further, the receivers that are not broadcast capable fetch the content via the bonded unicast capacity.

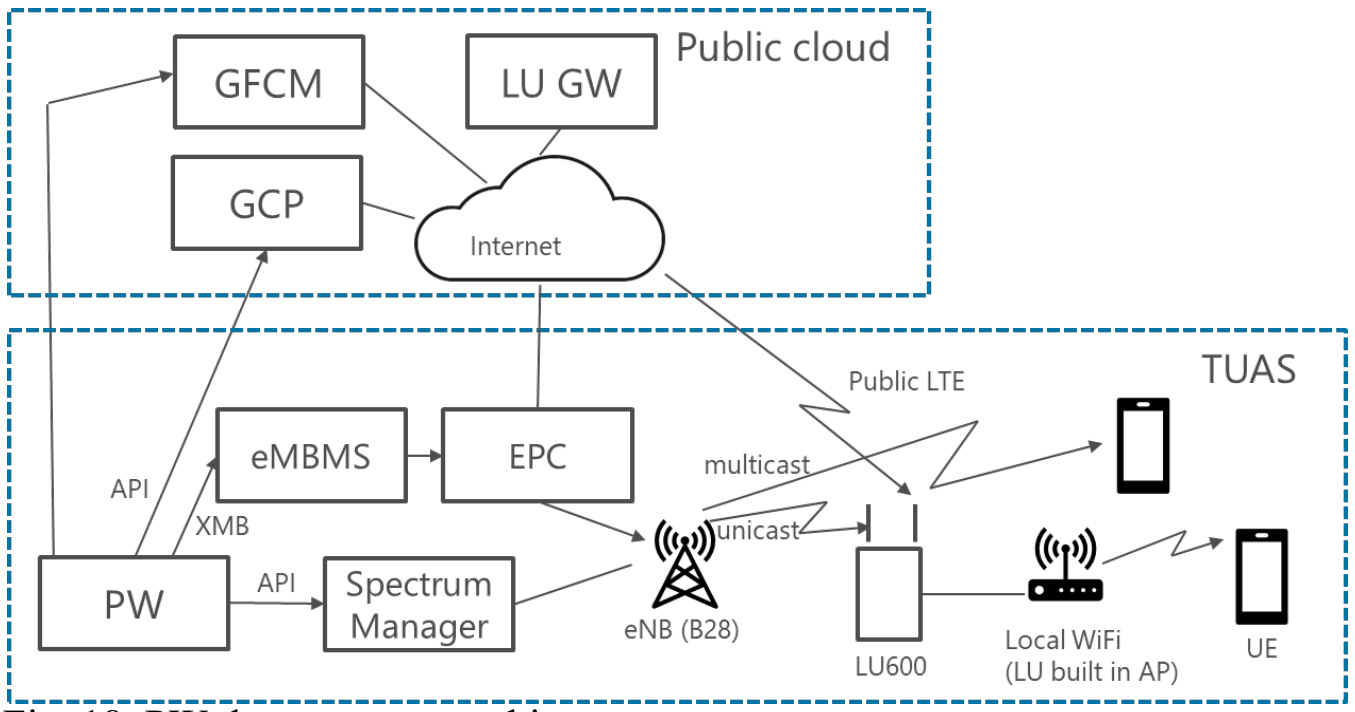

Fig. 18. PW demonstrator architecture.

As detailed in [20][34], the system has been successfully demonstrated, simulating an accident taking place in a port. Alert content describing the incident and providing instructions for the alert recipients has been generated and transmitted using the developed system.

The delivery using eMBMS appeared to be robust. Dynamic spectrum allocation has been demonstrated successfully. In the demonstration both LiveU multi-link Databridge and eMBMS devices connected rather quickly to the dynamically added network as soon as it became available. Bonding connections on the LiveU Databridge has been demonstrated with multiple LTE networks being used concurrently to deliver the data to several UEs, where the LiveU Databridge acted as a Wi-Fi access point via which the alerts were transmitted to the UEs.

## IX. Demonstrator: Reliable Multicast Delivery in 5G Networks

This demonstration shows the gains achieved by reliable multicast delivery in 5G networks. This includes the gains in and trade-offs among resource consumption, spectral efficiency, and QoE when multicast is introduced as a radio access network optimization - against unicast delivery mode - for delivering popular content. The demonstration also highlights the effects of using application layer methods, such as MPEG-DASH [35] streaming and multi-link delivery, on the efficiency and reliability of multicast delivery. Overall, the observations aim to contribute to a deeper understanding of the improvements achieved and to derive the conditions under which deployment of such technologies benefits the end-to-end network.

### A. *Concept and Relation to Use Cases*

In light of the abovementioned objective, this demonstrator focuses on the M&E vertical and uses PTM delivery for popular multimedia content, such as Olympic games or weekly-aired TV show where most users consume the same content. To maintain and/or provide high user experience when compared to pure unicast transmission, the demonstrator also shows the potential benefits of using a combination of multicast and unicast mode of deliveries for the UE that could be affected by the multicast service coverage, for instance at the edge of an MBMS area.

### B. *Technical Description*

The demonstration is carried out for different scenarios where the content delivery is done via 1) PTP only, 2) PTM

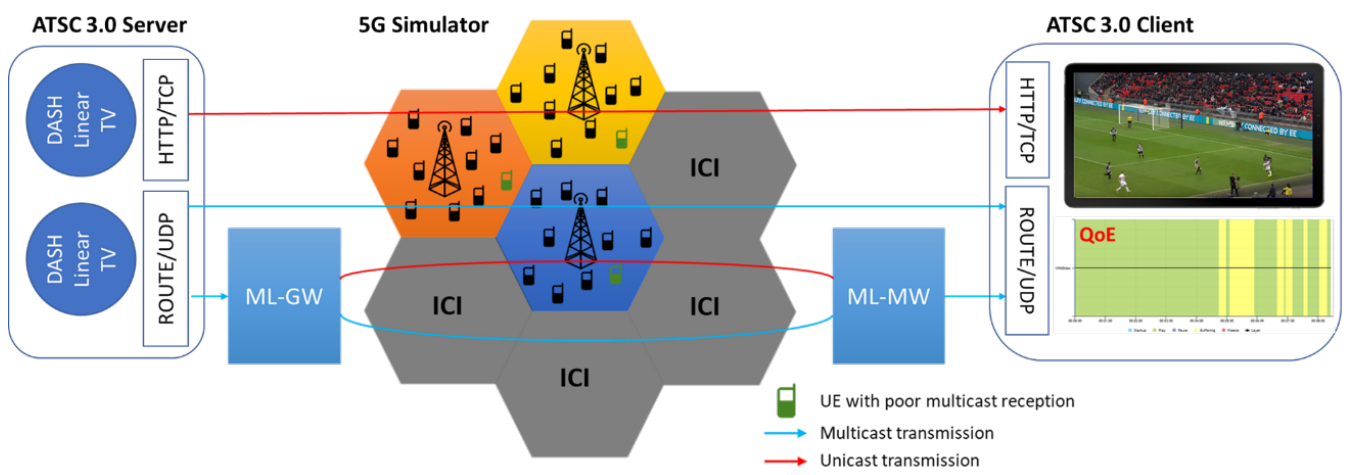

Fig. 19. Architectural framework of the demonstration of Reliable Multicast Delivery in 5G.

only, and 3) PTM with multi-link for the UE with poor multicast channel conditions.

Architectural framework of the demonstrator is shown in Fig. 19. The demonstration is comprised of an Advanced Television Systems Committee (ATSC) 3.0 service layer [36] content server, 5G system-level simulator and an ATSC 3.0 service layer client of which the roles are as follows:

- ATSC 3.0 content server is responsible for starting the service transmission. It hosts the content that is pre-encoded and packaged according to MPEG-DASH and serves it either upon request via HTTP/Transmission Control Protocol - TCP or linear via ROUTE/User Datagram Protocol - UDP to the 5G system-level simulator.
- 5G system-level simulator initiates the PTP and/or PTM communication and acts as a forwarding entity of the incoming media service. Also, it handles the multi-link functionality, where it is enabled based on the comparison between the effective signal to interference plus noise ratio (SINR) of the UE and pre-determined threshold. This solution includes a 1) ML-GW, where the multicast stream packets are duplicated onto a newly instantiated unicast session and a 2) ML-MW, where packets received from unicast and multicast are ordered and merged into one stream. Furthermore, to better demonstrate the scenarios, the simulator emulates additional UEs in the cells to receive similar DASH traffic generated by the simulator. Additionally, it provides real-time monitoring of network-related key performance indicators (KPIs) and QoE through its GUI.
- ATSC 3.0 client as an Android tablet receives the content either via unicast (HTTP/TCP) or multicast (ROUTE/UDP) which is then consumed for playback. This client is also responsible for collecting related QoE metrics for real-time QoE monitoring and reporting service.

### C. Demonstration Setup and Results

Based on the provided description and architecture, ATSC 3.0 service layer software for both server and client are interfaced with the 5G system-level simulator provided by Nomor Research. The multimedia content given by BT is prepared and stored on the server. Multi-Link logic described previously is provided by BLB and integrated into the simulator. MPEG-DASH compatible ExoPlayer is used for content playback, which is also interfaced with the Smartlib QoE Library and QoE Analytics solutions from Broadpeak for QoE monitoring and reporting. 5G network simulation environment and the used content parameters are provisioned as in Table 1. The demonstrator setup is as provided in Fig. 20.

TABLE 1. DEMONSTRATION SETUP PARAMETERS

| | Parameter | Value |
|---|---|---|
| **Network** | Carrier frequency | 3.5GHz |
| | Total BS transmit power | 51dBm |
| | System bandwidth | 100MHz |
| | BS antenna configuration | $[M, N, P] = [8, 4, 2]$[1] |
| | BS TXRU configuration | $[M_P, N_P, P] = [1, 4, 2]$[2] |
| | UE antenna / TXRU configuration | $[M, N, P] = [1, 4, 2]$ |
| | UE mobility model | 3kmph, randomly uniform distr. |
| | BS noise figure | 5dB |
| | UE noise figure | 9dB |
| | BS ant. element gain | 5dBi |
| | BS ant. elevation – 3dB-BW | 65° |
| | BS ant. azimuth – 3dB-BW | 65° |
| | BS ant. elec. downtilt | 20° |
| | Deployment type | Urban |
| | Network layout | Homogeneous macro cells |
| | Number of cells | 3 simulated cells wrapped by surrounding cells that simulate inter-cell interference |
| | Inter-site distance | 200m |
| | UE density per cell | 10 |
| | Multicast MCS index w/o and w/ multi-link[3] | 2, 4 |
| | Multi-Link Switching Threshold | 5dB |
| | Traffic type | DASH<br>Adaptation algorithm [37]<br>Segment download cancellation[4] [38]<br>Session quit timer of 30s[5] |
| **Content** | Unicast content encoding | 1Mbps@480p, 4Mbps@1080p, 8Mbps@1080p, 12Mbps@2K, 16Mbps@4K, 20Mbps@4K<br>24 fps<br>HEVC |
| | Multicast content encoding | 20Mbps@4K<br>24 fps<br>HEVC |

As aforementioned, KPIs in this work are resource consumption, spectral efficiency and QoE which are defined as follows:

- *Average resource consumption:* It is the average percentage of the resources used up for the service delivery

[1] $M$, $N$ and $P$ refer to the number of vertical, horizontal and polarization arrangement of antenna elements, respectively.

[2] The 8 vertical antenna elements for each polarization are hard-wired and are connected to a TXRU.

[3] Multicast MCS index is selected to provide as low error rate as possible while maintaining sufficient transmission rate for the service delivery.

[4] Segment download cancellation is integrated for further robustness in the content playout to take into account the sudden network condition changes of UE.

[5] Session quit logic is integrated to emulate a representative user behavior, in which the UE stops streaming the content when the DASH segment reception takes too long (30s) and segment download cancellation did not change this situation.

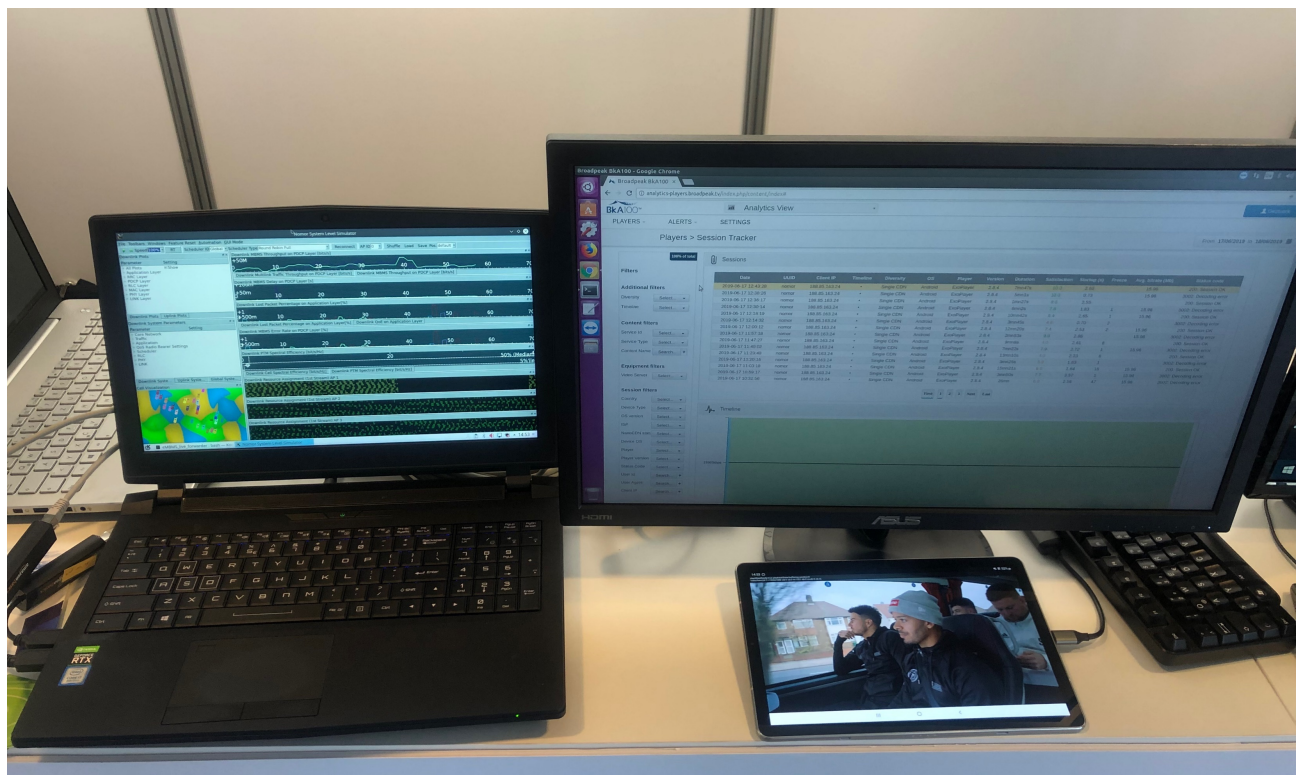

Fig. 20. Demonstration setup: ATSC.3.0 content server (left-back PC), 5G system-level simulator (left-front PC), ATSC 3.0 client (Android tablet), QoE Analytics server (right PC).

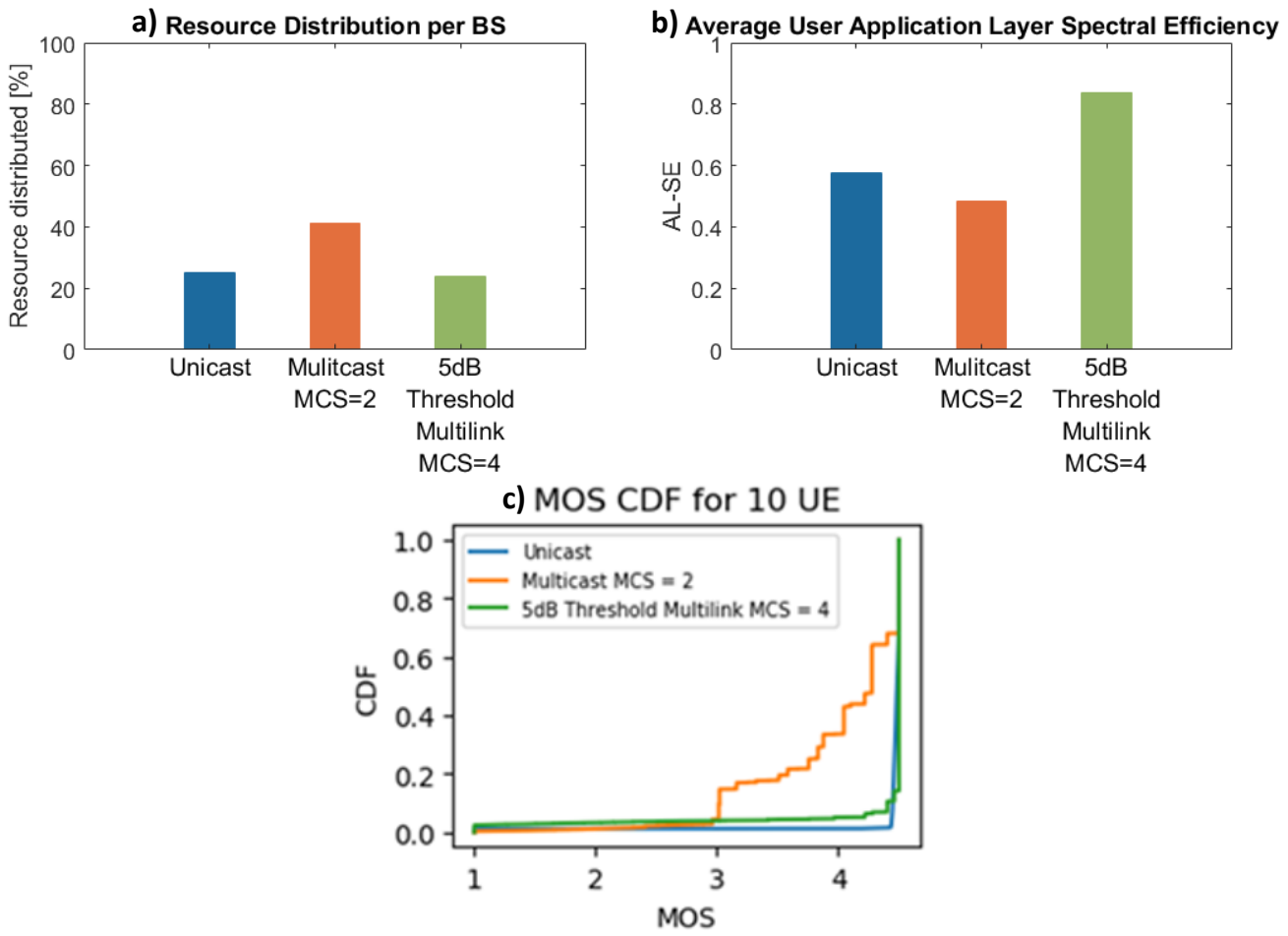

Fig. 21. Observed results from the demonstration: a) Average resource consumption, b) Average user application layer spectral efficiency, c) QoE CDF.

and calculated as the ratio of the average number of physical resource blocks used and the total number of available radio resources.

- *Average user application layer spectral efficiency (AL-SE):* It is the average user spectral efficiency from the perspective of the application layer and measured as the ratio of the total number of bits generated by the content source and the average bandwidth usage which is system bandwidth multiplied by the average resource consumption. It is noted that packet losses caused by transmission errors or congestion are not considered in this KPI which is why the interpretation of this must be done along with QoE KPI into which the packet losses are reflected.
- *QoE:* The quantitative estimation of the subjective user experience, i.e., Mean Opinion Score (MOS), is calculated for each UE. For the ATSC 3.0 client, the QoE library provided from Broadpeak is used. For the emulated UEs within the simulator, the work in [39] is integrated into the simulator. As an extension to [39], for observing results in real-time, the QoE score is computed in a sliding window fashion with a window size of 15 MPEG-DASH segments.

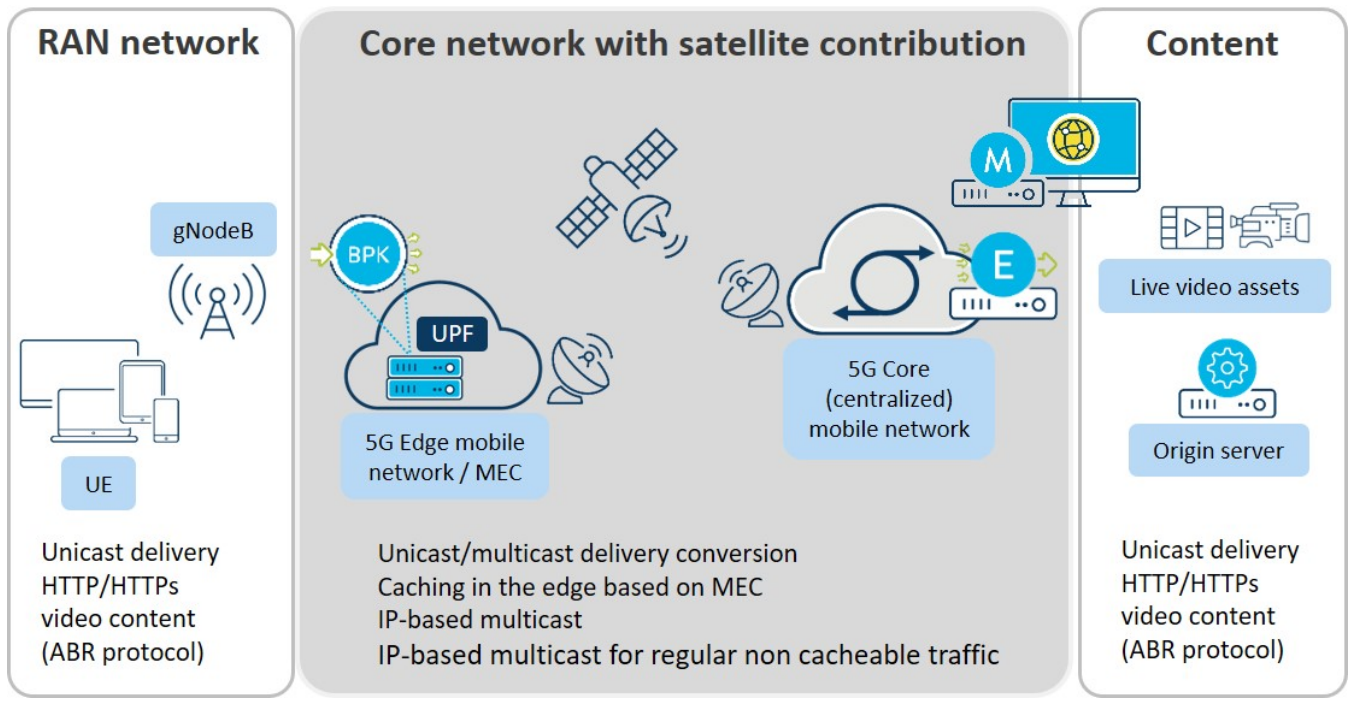

Fig. 22. Demonstrator concept for Over-the-Air Multicast over Satellite for Video Caching and Live Content Delivery.

In this way, a forgetting factor in user behaviour, where the good or bad user experience is forgotten over time, is also accounted for.

The observed results from this demonstration setup are provided in Fig. 21. From these, it is seen that multicast can be used for optimizing the resource usage. Compared to pure multicast, multi-link with a density of 10 UEs per cell and 5dB multi-link switching threshold proves to be beneficial for increase in service quality for a larger portion of the UEs. Also, in multi-link scenarios, the perfect service quality is observed for approximately 90% of UEs as compared to unicast, where approximately 95% coverage is achieved. Additionally, multi-link achieved an AL-SE that is approximately 50% higher than that of the unicast configuration.

However, it should be emphasized that the benefit of the multi-link logic used specifically in this demonstration could be limited for larger user densities as it completely duplicates the delivered stream.

## X. Joint Demonstrator with SAT5G: Optimized Resources Allocation For Live Video Content

This is a joint demonstrator developed by 5G-Xcast & SaT5G 5G-PPP projects involving Avanti's high throughput HYLAS 4 geostationary orbit (GEO) satellite capacity, Broadpeak's MEC-enabled platform for the content delivery network, University of Surrey's 5GIC testbed network and VT iDirect's 5G enabled satellite hub platform and satellite terminals [20][40]. This demonstrator focuses on using satellite multicast capabilities to deliver live channels to a 5G edge mobile network. The main objectives of the demonstrator are:

- To improve video distribution efficiency using multicast Adaptive Bitrate (mABR) over Satellite as contribution link,
- To minimize end-to-end latency using Common Media Application Format (CMAF) combined with chunked transfer encoding (CTE) DASH over mABR link,
- To address all screens thanks to transparent use of local cache servers,
- To Provide synchronized video delivery on any screen.

In this paper, we briefly introduce the concept of this demonstrator, that is to showcase over-the-air satellite multicast technology for the delivery of live channels using a MEC platform for the content delivery network (CDN) integration with efficient edge content delivery, as illustrated in Fig. 22. This demonstrator highlights the benefits in terms of bandwidth efficiency and delivery cost of using a satellite-enabled link for provisioning live content in a 5G system.

## XI. Discussions

The demonstrations shown in this paper illustrate some of the innovative technical solutions that 5G potentially enables, including deployment of 5G in the future of media delivery. The following discussions on the key findings of these demonstrators and showcase suggest that new opportunities or challenges for broadcasters and network operators might arise in future media delivery.

Hybrid Broadcast Service with linear TV and add-on content, 5G-Xcast Showcase in European Championships 2018: The HBS demonstrator and showcase illuminates the potential of PTM capabilities in 5G, especially for large-scale distribution of national or even world-wide popular content. The live signal was received, free-to-air, on both an IPTV set and commercially available smartphones at the same time. Smartphones were equipped with the eMBMS functionality and the app that enables tuning to different TV channels available in the broadcast stream. In addition to the TV channels delivered via the broadcast component of the LTE signal, on-demand content was provided over the unicast connection. For this, the HbbTV standard was used on both the smartphones and the TV sets. On the former the on-demand content could be accessed via the app, on the latter using the "red button".

Hybrid Broadcast Service with MooD: The demonstrator has shown a unified framework, in this case a 5G network, being able to convey different media services according to operator and user demands. From traditional always-on linear TV services to on-demand streaming; from user-agnostic delivery, to adaptive streaming, to device capabilities; from fixed TV receivers to users on the move. All within the same system.

Converged, autonomous MooD in fixed and mobile networks: This demonstrator has shown how multicast/broadcast can be used as an internal network optimization for applications that are unmodified. This is an advance over current systems such as eMBMS that require MBMS-aware applications and extensive service-specific configuration. The approach taken here used the existing unicast interface with content service providers and can be applicable to both fixed and mobile networks.

Object-based Media across 5G networks using Dynamic Adaptive Streaming over IP Multicast: In this demonstrator, we have shown that DASM system is used to ensure the reliable delivery of multicast objects which are commonly used and/or bandwidth-heavy, whilst the less commonly used and/or less bandwidth-expensive objects are delivered over unicast HTTP. A hybrid of multicast and unicast objects being rendered seamlessly on the user's device can potentially provide an immersive user experience.

Hybrid Broadcast Service with Multi-Link: Here we have demonstrated the repair mechanism and the transmission over different physical interfaces. The analysis of the captured traffic on the three sites (Server, Client and eNB) showed that the repair mechanism is activated when the video stream over LTE eMBMS has problems and therefore packets can be requested using unicast even over another interface. Note that enough buffering is necessary at the receiver side in order to prevent disruption in the visualization of the content. In this setup, only one client is involved in the demonstration. In further developments, more clients can receive a live stream in a synchronized manner.

Multimedia Public Warning and Spectrum Management demonstrator: Using eMBMS we have demonstrated an efficient and robust delivery of the multimedia public warning content to end-users. This enables the relevant emergency authorities to send significantly richer information. Addition of dynamic spectrum utilization and bonded connections further enhances the robustness and the ability of the users to receive alerts with different types of terminals.

Reliable multicast delivery in 5G networks: This demonstrator has revealed the potential of the multicast mode of delivery in 5G networks and how it can be enhanced in terms of efficiency, reliability as well as end user experience by using application layer techniques like multi-link and MPEG-DASH.